\documentclass[lettersize,journal]{IEEEtran}
\usepackage{amsmath,amsfonts}
\usepackage{algorithmic}
\usepackage{algorithm}
\usepackage{array}
\usepackage[table,xcdraw]{xcolor}
\usepackage[caption=false,font=normalsize,labelfont=sf,textfont=sf]{subfig}
\usepackage{textcomp}
\usepackage{stfloats}
\usepackage{url}
\usepackage{verbatim}
\usepackage{graphicx}
\usepackage{cite}
\usepackage{nomencl}
\makenomenclature

\begin{document}

\title{Aggregated distribution grid flexibilities in subtransmission grid operational management}

\author{Neelotpal Majumdar,~\IEEEmembership{student member,~IEEE,} and Lutz Hofmann
\thanks{The study performed under the research project "SiNED - System Services for secure electricity grids in times of advancing energy transition and digital transformation" acknowledges the support of the Lower Saxony Ministry of Science and Culture through the "Niedersächsisches Vorab" grant program (grant ZN3563) and of the Energy Research Centre of Lower Saxony.''}
\thanks{Neelotpal Majumdar and Lutz Hofmann are associated with the Electric Power Engineering Section, Institute of Electric Power Systems at the Leibniz University of Hannover, Appelstrasse 9A, 30167 Hannover, Germany (email: majumdar@ifes.uni-hannover.de (Neelotpal Majumdar), hofmann@ifes.uni-hannover.de (Lutz Hofmann))}
\thanks{Manuscript received Month mm, yyyy; revised August mm, yyyy.}}

\markboth{Journal of \LaTeX\ Class Files,~Vol.~14, No.~8, August~2021}%
{Shell \MakeLowercase{\textit{et al.}}: A Sample Article Using IEEEtran.cls for IEEE Journals}

\IEEEpubid{0000--0000/00\$00.00~\copyright~2023 IEEE}

\maketitle

\begin{abstract}
Aggregated flexibilities or PQ-capabilities (active and reactive power capabilities) are termed in literature as Feasible Operating Regions (FORs). The FORs from underlying active distribution grids can effectively contribute to the operational management at the HV grid level. The HV buses are allocated aggregated FORs from the underlying MV grids, which are inherently nonlinear and non-convex. Therefore, two approaches are proposed in the paper to apply the FOR constraints in the HV grid operational management. First, a mixed integer linear programming (MILP) based optimization approach for alleviating the HV grid constraint violations is proposed, which addresses the non-convexity of the FOR using piecewise segmentation. Furthermore, the MILP method is enhanced to consider the influence of the HV bus voltage on the underlying MV grid flexibilities resulting in a three dimensional PQ(V)-FOR. Second, a convexification approach is proposed, which uses a convex approximation of the non-convex 3D PQ(V)-FOR shape for implementation in a linear optimization method. Results reveal a robust utilization of the distribution flexibilities to maintain grid security and reliability at the HV grid level. Comparisons present increased computation times for the MILP method which are significantly improved using the convexification based approach.
\end{abstract}


\IEEEpubidadjcol 

\nomenclature[01]{\(\text{LV}\)}{Low voltage grid level}
\nomenclature[02]{\(\text{MV}\)}{Medium voltage grid level}
\nomenclature[03]{\(\text{HV}\)}{High voltage grid level}
\nomenclature[04]{\(\text{EHV}\)}{Extra high voltage grid level}
\nomenclature[05]{\(\text{FOR}\)}{Feasible operating region}
\nomenclature[06]{\(\text{TSO}\)}{Transmission system operator}
\nomenclature[07]{\(\text{DSO}\)}{Distribution system operator}
\nomenclature[08]{\(\text{MILP}\)}{Mixed integer linear programming}
\nomenclature[09]{\(\text{PQ}\)}{Active and reactive power flexibility}
\nomenclature[10]{\(\text{PQ(V)}\)}{Active and reactive power flexibility subject to specified slack voltages at the overlaying HV bus}
\nomenclature[11]{\(\text{DER}\)}{Distributed energy resources}
\nomenclature[12]{\(p_i\)}{Bus active power injection at bus index $i$}
\nomenclature[13]{\(q_i\)}{Bus reactive power injection at bus index $i$}
\nomenclature[14]{\(v_i\)}{Bus voltage at bus index $i$}
\nomenclature[15]{\(\delta_i\)}{Bus voltage angle at bus index $i$}
\nomenclature[16]{\(y_{ii}\)}{Bus shunt admittance at bus index $i$}
\nomenclature[17]{\(y_{ij}\)}{Branch admittance between bus indices $i$ and $j$}
\nomenclature[18]{\(\theta_{ii}\)}{Admittance phase angle at bus index $i$}
\nomenclature[19]{\(\theta_{ij}\)}{Branch admittance phase angle between bus indices $i$ and $j$}
\nomenclature[20]{\(\boldsymbol{ID}_{\text{TT}}\)}{Terminal current to terminal bus voltage angles sensitivity matrix}
\nomenclature[21]{\(\boldsymbol{IU}_{\text{TT}}\)}{Terminal current to terminal bus voltage magnitude sensitivity matrix}
\nomenclature[22]{\(P_\mathrm{vert}\)}{Vertical interconnection active power flow}
\nomenclature[23]{\(Q_\mathrm{vert}\)}{Vertical interconnection reactive power flow}
\nomenclature[24]{\(\Delta\boldsymbol{p}\)}{Vector of bus (nodal) active power flexibility injections as optimization variables}
\nomenclature[25]{\(\Delta\boldsymbol{q}\)}{Vector of bus (nodal) reactive power flexibility injections as optimization variables}
\nomenclature[26]{\(\Delta\boldsymbol{v}\)}{Vector of bus (nodal) voltages as optimization variables}
\nomenclature[27]{\(\Delta\boldsymbol{i}\)}{Vector of branch terminal currents as optimization variables}

\printnomenclature
\IEEEpubidadjcol 
\section{Introduction}
\IEEEPARstart{P}{hasing} out of thermal power plants requires the ever expanding renewable energy sector to ensure a secure power supply. Active and reactive power reserves for ensuring a reliable grid operation require to be supplied from the renewables, the major share of which are installed at the distribution grid level \cite{wussow2021sined}. Therefore, active distribution networks are assuming a primary role for ancillary services provision at the local grid level and for the overlaying sub-transmission and transmission grids \cite{lotz2021potentials}. Researches investigate FOR determination techniques to quantify the aggregate PQ-flexibility potential available from the underlying distribution grid levels. Different methods using stochastics, mathematical optimization and heuristic based approaches are proposed \cite{heleno2015estimation}\nocite{ageeva2019analysis}\nocite{gonzalez2018determination}\nocite{riaz2019feasibility}\nocite{silva2018estimating}\nocite{rossi2017fast}\nocite{contreras2018improved}\nocite{contreras2019impact} -\cite{sarstedt2021survey}. An approach for aggregation assuring robustness considering uncertainties and N-1 contingencies based on deep reinforcement learning is further proposed in \cite{wang2022robust}. The FORs determined are aggregated at single interconnections and represent the flexibility potential from the underlying MV and LV grids. The aggregated PQ-flexibility potentials serve as a reliable interface for coordination between transmission system operators (TSO) and distribution system operators (DSO). The TSOs can integrate the FORs in grid planning and efficient operational management. A study presents the application of the aggregated distribution grid FORs for addressing congestion problems at the overlaying HV grid levels \cite{contreras2021congestion}. Results reveal the practical usage of distribution flexibilities for ancillary services provision in multiple voltage level context. The undertaken study improves upon the concept by additionally considering steady stage voltage limit violations. Corrective PQ-flexibility procurement from the aggregated FORs at the HV buses along  with local flexibilities e.g. HV wind parks are determined using a linear optimization algorithm. Furthermore, the aggregated FORs are inherently nonlinear and non-convex owing to the adherence to distribution grid constraints and generating device constraints. Therefore, in the undertaken research a mixed integer linear programming (MILP) based method and a convex approximation based linear programming are proposed for considering the non-convex FORs. The MILP based strategy provides efficient utilization of the non-convex FORs and obtains optimum solutions for the aforementioned corrective actions. The convexification based approach further improves the computation times significantly while ensuring permissible and optimum solutions. The methods are developed considering the effect of the HV bus voltage from the overlaying grid on the aggregated PQ-flexibility potential of the underlying grids. Therefore, 3D PQ(V)-FORs are aggregated from the underlying MV grid levels, introduced for the first time in \cite{schwerdfeger2017vertikaler} and subsequently discussed in \cite{sarstedt2020simulation}. The corresponding application for the HV grid operational management is presented in \cite{schwerdfeger2017vertikaler} using a Tabu-search algorithm. The method, however, considers a 2D FOR as a subset of the entire volume, that is valid for the entire range of the HV bus voltage. A particle swarm optimization method for solving a HV grid congestion using an aggregated 3D PQ(V)-FOR at a single HV bus is presented in \cite{sarstedt2023Technoökonomische}. However, this results in increased computation times for acquiring flexibilities from a single bus, rendering it impracticable for short term operational planning. The method further uses a simplified interpolation to traverse the voltage range of the 3D PQ(V)-FOR, and limits flexible usage of the polyhedral volume. The concept is enhanced in this paper, using mathematical optimization techniques to improve computation times and flexible usage of the volume of the 3D PQ(V)-FORs. Furthermore, scalability is demonstrated considering multiple 3D PQ(V)-FORs at multiple HV buses in the HV grid operational management. 

The paper is subdivided into the following sections: section II presents the concept of FOR application for mitigating technical grid problems at overlaying voltage levels; section III introduces the mathematical formulations used for the MILP optimization based on a simplified 2D FOR; section IV discusses the enhanced 3D PQ(V)-FOR considering the HV bus voltage as a 3rd dimension and the modified MILP method for consideration of the 3D FOR shape; section V presents the convex approximation method of the 3D PQ(V)-FOR for utilization in a linear programming environment; section VI presents the investigated grid HV model and the results demonstrating efficacy of the methods for investigated scenarios and section VII discusses the conclusion with further interesting research directions.

\section{Concept: Ancillary services provision at HV grid through aggregated distribution grid FORs}

\begin{figure}
    \centering
    \includegraphics[width=9cm]{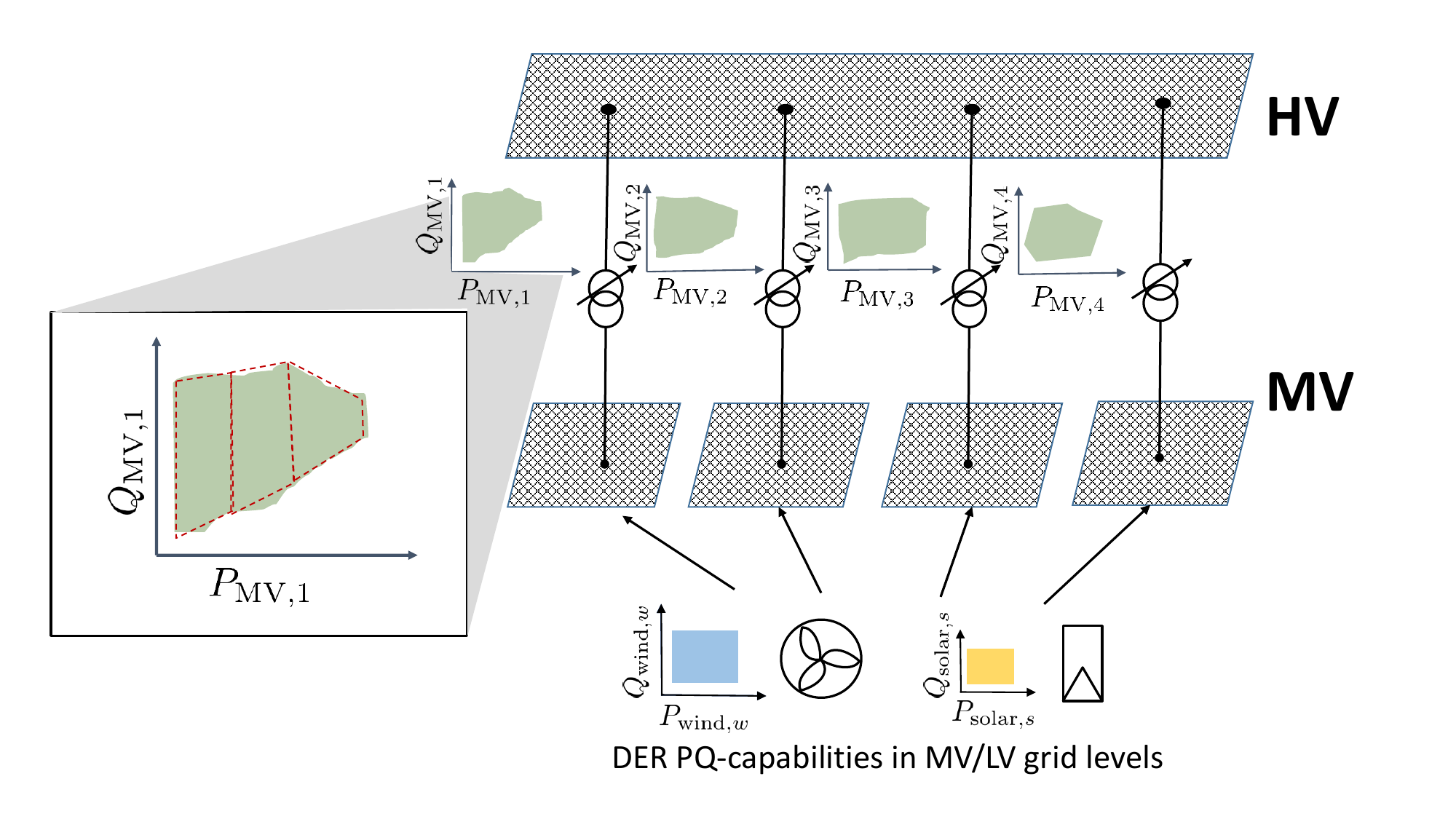}{}
    \caption{Schematic diagram representing aggregated FOR utilization for mitigating steady state HV grid problems.}
    \label{fig:concept}
\end{figure}

The distribution grid aggregated FORs present additional PQ-flexibilities at the HV buses complementing the installed flexibilities e.g. DERs, conventional generators and FACTS devices in the HV system. Preventive or curative measures in response to prognosed or transpired grid constraint violations can be addressed using the available HV bus flexibilities. The undertaken study is limited to steady state voltage limit violations and congestion problems. The strategy is applicable both for preventive as well as curative measures, depending on the grid requirements. Fig. \ref{fig:concept} presents a conceptual understanding of addressing HV grid problems using aggregated distribution grid flexibilities. Typical simplified rectangular PQ-capability curves are assumed for wind and photovoltaic DERs, according to specifications mentioned in \cite{majumdar2022linear}. The other variants for the PQ-capability curves can be analogously considered, e.g. the triangular or circular variations. However, these variations are not considered in the scope of the undertaken study.

The DER flexibilities are presented as PQ-capability curves for the wind and photovoltaic DERs at the buses of the MV/LV grids in the schematic representation. Subsequently, the PQ-flexibility potentials are aggregated at the corresponding HV/MV interconnections using the optimal power flow framework presented in \cite{majumdar2022linear}. In the schematic representation aggregated FORs for the four exemplary MV grids $\text{MV},$1-4 are demonstrated. The aggregated FORs determind are nonlinear and non-convex which is unexpected considering the linear PQ-capability curves of the constituent DERs. However, the aggregation process is subject to technical grid constraints (e.g. bus voltage limits, thermal current limits), grid power losses and generator power constraints, thus, producing the resultant shapes. Power system operational management at the HV grid level requires efficient operations of the grid flexibilities e.g. aggregated FORs, local HV grid DERs and FACTs devices. Two approaches are presented in this paper, which considers the HV grid flexibilities  for solving steady state voltage limit violations and congestions. Owing to the non-convexity of the aggregated FORs a corresponding piecewise linearization approach of the FORs is proposed using a MILP based method. Fig. \ref{fig:concept} presents the piecewise linearization of a representative FOR determined from the aggregated PQ-flexibilities of the $\text{MV},1$ grid. The non-convex area is segmented into three convex areas formulated as a linear system of equations. The formulation is correspondingly integrated into the MILP optimization environment. The mixed integer formulation is required to optimally utilize the specific segment as required for the mitigation of the grid constraint violations.

\section{Optimization algorithm for the operational management of the HV grid}
This section subdivides the mathematical formulation involved in the optimization algorithm. At first, the power flow sensitivities are presented, followed by the segmentation of the power injections and subsequently the corresponding integration into the optimization environment. The mathematical syntax used throughout the paper is as follows:
\begin{itemize}
    \item The matrices are presented in bold, italic and uppercase e.g. $\boldsymbol{A}$
    \item The one dimensional vectors are presented in bold, italic and lowercase e.g. $\boldsymbol{a}$
    \item Variables are represented by lowercase and italic e.g. $a$
\end{itemize}

\subsection{Power flow sensitivities}

The inherently nonlinear and non-convex power flow equations are linearized using the Jacobian matrix derived from the Newton-Raphson load flow equations presented in \cite{majumdar2022linear}. The bus active and reactive power injection are accordingly a function of bus voltages and grid admittances.
\begin{equation}p_i = v_i^2y_{ii}\cos\theta_{ii}+\sum_{\substack{j=1 \\i\neq j}}^{n}v_iv_jy_{ij}\cos(\delta_i-\delta_j-\theta_{ij})\label{eq:newtonP}\end{equation}
\begin{equation}q_i = -v_i^2y_{ii}\sin\theta_{ii}+\sum_{\substack{j=1 \\i\neq j}}^{n}v_iv_jy_{ij}\sin(\delta_i-\delta_j-\theta_{ij})\label{eq:newtonQ}\end{equation}
where $i,j \in n$ represent the bus or node indices \footnote{in the applied formulation however, the shunt power flows are not modelled as a function of shunt admittances and bus voltages, but as constant power sources/loads, such that $v_i^2y_{ii}\cos\theta_{ii}=p_{i,\mathrm{const}}$, $-v_i^2y_{ii}\sin\theta_{ii}=q_{i,\mathrm{const}}$}. The corresponding inverse Jacobian $\boldsymbol{J}^{-1}$ describes the bus voltage magnitude and angle sensitivities in response to power deviations, as discussed in \cite{majumdar2022linear}.
\begin{equation}
\begin{bmatrix}
	\Delta \boldsymbol{\delta} \\
	\Delta \boldsymbol{v}
\end{bmatrix}=\boldsymbol{J}^{-1}\begin{bmatrix}
\Delta \boldsymbol{p} \\
\Delta \boldsymbol{q}
\end{bmatrix}\label{eq:jacobianInv}\end{equation}
where $\Delta \boldsymbol{p}$ and $\Delta \boldsymbol{q}$ represent the vector of deviations in bus active and reactive power injections. Subsequently, the resultant deviations in bus voltage magnitudes $\Delta \boldsymbol{v}$ and angles $\Delta \boldsymbol{\delta}$ are required for determining branch current sensitivities to branch terminal voltages and angles, according to an established formulation presented in \cite{leveringhaus2014comparison}.
\begin{equation}
\Delta \boldsymbol{i} = \boldsymbol{ID}_\text{TT}\Delta\boldsymbol{\delta}_\text{T} + \boldsymbol{IU}_\text{TT}\Delta\boldsymbol{v}_\text{T}
\label{eq:7}\end{equation}
The subscript TT describes the branch terminal to branch terminal indices represented by bus indices $i,j$. A single subscript T correspondingly refers to the branch terminals. The terminal sensitivity matrices $\boldsymbol{ID}_\text{TT}$ and $\boldsymbol{IU}_\text{TT}$ are correspondingly multiplied by the bus to branch terminal incidence matrix. Thus, sensitivities with regards to bus voltages and angles are derived, for integration into the optimization environment.

The bus voltage and current deviations are subject to maximum and minimum technical grid constraints. Therefore, adherence to steady state bus voltage constraints and thermal branch current capacity is addressed.

\subsection{Segmented piecewise linearization of the aggregated FORs}

\begin{figure}
    \centering
    \includegraphics[width=8cm]{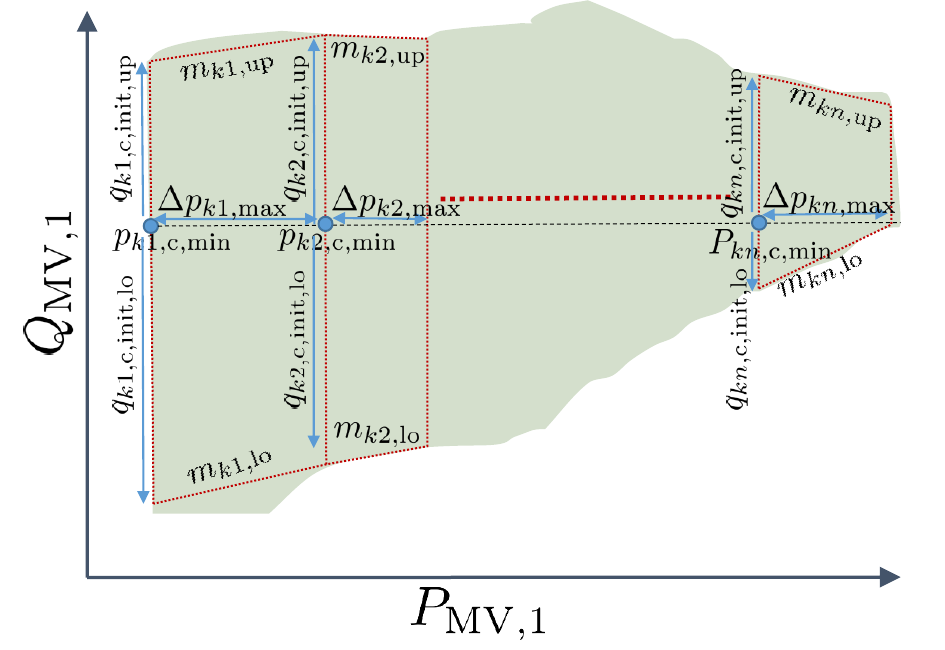}{}
    \caption{Linearization parameters for a schematic FOR segment}
    \label{fig:segment}
\end{figure}

The FORs are presented as a two dimensional PQ-flexibilitity aggregated from the underlying MV grid levels at the respective HV buses. An initial point of operation for each FOR represents the power demanded/injected by the underlying MV grid levels at the respective HV bus. The FORs are segmented into $ki_\text{max}$ sections as demonstrated in Figure \ref{fig:concept}. Each section is expressed in the form of linear equations for integration into the linear optimization framework. The adapted operating point in the FOR is selected from a specific section as the optimizer solution. This is enabled using the mixed integer formulation and is required to prevent multiple operating point adaptations within a particular FOR.
\begin{equation}
    \Delta p_i = \sum_{ki=1}^{ki_\text{max}}\Delta p_{ki} + \sum_{ki=1}^{ki_\text{max}} x_{ki} \cdot p_{ki,\text{c,min}}
    \label{eq:psegment1}
\end{equation}
\begin{equation}
    0 \leq \Delta p_{ki} \leq x_{ki} \cdot \Delta p_{ki,\text{max}}
    \label{eq:psegment2}
\end{equation}
Therefore, the active power flexibility (P-flexibility) $\Delta p_i$ with index '$i$' is activated when the integer variable $x_{ki} \neq 0$. Furthermore, the segment corresponding to the index $ki$ is selected when $x_{ki} = 1$, where $\Delta p_{ki}$ refers to the P-segment and $P_{ki,\text{c,min}}$ refers to a constant minimum or intercept. $\Delta p_{ki,\text{max}}$ is a constant term and imposes an upper limit on the variable $\Delta p_{ki}$. The selection of a single segment is ensured using an integer formulation as presented.
\begin{equation}
    \sum_{ki=1}^{ki_\text{max}} x_{ki}=1, \quad x_{ki}=[0,1]
    \label{eq:integer}
\end{equation}
Thus, $\Delta p_{ki} = [0,\Delta p_{ki,\text{max}}]$ is bounded corresponding to the limits imposed by a multiplication with the integer variable $x_{ki}=[0,1]$. Accordingly, for a particular $x_{ki}=1$, eq. (\ref{eq:psegment1}) and (\ref{eq:psegment2}) yield the effective change in bus power injection. 
\begin{equation}
    \Delta p_i = \Delta p_{ki} + p_{ki,\text{c,min}}
\end{equation}
Subsequently, the reactive power flexibility (Q-flexibility) is required to be guaranteed from the selected segment '$ki$'. 
\begin{equation}
    \Delta q_i \leq \sum_{ki=1}^{ki_\text{max}}m_{ki,\text{up}}\Delta p_{ki} + \sum_{ki=1}^{ki_\text{max}} x_{ki} \cdot q_{ki,\text{c,init,up}}
    \label{eq:qsegmentup}
\end{equation}
\begin{equation}
    \Delta q_i \geq \sum_{ki=1}^{ki_\text{max}}m_{ki,\text{lo}}\Delta p_{ki} + \sum_{ki=1}^{ki_\text{max}} x_{ki} \cdot q_{ki,\text{c,init,lo}}
    \label{eq:qsegmentlow}
\end{equation}
The slope $m_{ki}=\frac{q_{ki+1,\text{c,init}}-q_{ki,\text{c,init}}}{\Delta p_{ki,\text{max}}}$ is segregated for the upper and lower edges of the segment, indexed by 'up' and 'lo' respectively. $q_{ki,\text{c,init,up}}$ and $q_{ki,\text{c,init,lo}}$ represent constant initial values for the upper and lower edges of the segment respectively. A detailed description of the parameters introduced in eq. (\ref{eq:psegment1})-(\ref{eq:qsegmentlow}), is graphically represented in Fig. \ref{fig:segment}. For clarity, the piecewise segmentation is demonstrated with the schematic representation of the FOR from the '$\mathrm{MV,1}$' grid  presented in Fig. \ref{fig:concept}. The trapezoidal convex segments are depicted by 'red' dotted lines. The upper and lower segments are separated, based on an initial operation point of the grid, marked by the 'blue' circles. 
The optimization problem can accordingly be formulated.
\begin{equation}
\renewcommand*{\arraystretch}{1.5}
\begin{gathered}
\begin{split} \text{min} \quad & \boldsymbol{c}^\text{T}\boldsymbol{x} \mid \boldsymbol{x} = [\Delta \boldsymbol{p}^\text{T}, \Delta \boldsymbol{q}^\text{T}, \Delta \boldsymbol{\delta}^\text{T}, \Delta \boldsymbol{v}^\text{T}, \boldsymbol{x}_s^\text{T}, \boldsymbol{x}_c^\text{T}, \boldsymbol{x}_\text{int}^\text{T}]^\text{T};\\\end{split}
\\ \begin{split}\text{s.t} \quad & \mathrm{eq. \ (\ref{eq:psegment1})-(\ref{eq:integer}),(\ref{eq:qsegmentup}),(\ref{eq:qsegmentlow})}\\
& \boldsymbol{v}_{\text{min}} \leq \boldsymbol{v}_0 + \Delta \boldsymbol{v} \leq \boldsymbol{v}_{\text{max}} \\ & \boldsymbol{i}_{\text{min}} \leq \boldsymbol{i}_0 + \Delta \boldsymbol{i} \leq \boldsymbol{i}_{\text{max}}\\\end{split}
\end{gathered}\label{eq:opt}\end{equation}
The dimension of the grid state vectors consists of 4 times the number of buses $n$, representing the grid states of bus active and reactive power injections, bus voltage angles and magnitudes. The slack bus is not considered in the linear optimization formulation, and corresponding rows and columns are removed from the sensitivity matrices. The variable vector $\boldsymbol{x}_s$ represents the vector of active and reactive power segments for every bus $i$ represented by the variable $\Delta p_{ki}$ and $\Delta q_{ki}$ respectively. The vector $\boldsymbol{x}_c$ represents the vector of constant initial terms for the active and reactive power segments, $p_{ki,\text{c,init}}$, $q_{ki,\text{c,init,up}}$ and $q_{ki,\text{c,init,lo}}$. Correspondingly, the vector term $\boldsymbol{x}_\text{int}$ depicts the vector set of integer variables $x_{ki}$ for the discrete active and reactive power segments.

\section{The three dimensional PQ(V)-FOR and adaptation of the MILP formulation}
\subsection{Concept of the three dimensional PQ(V)-FOR}
\begin{figure}
    \centering
    \includegraphics[width=9cm]{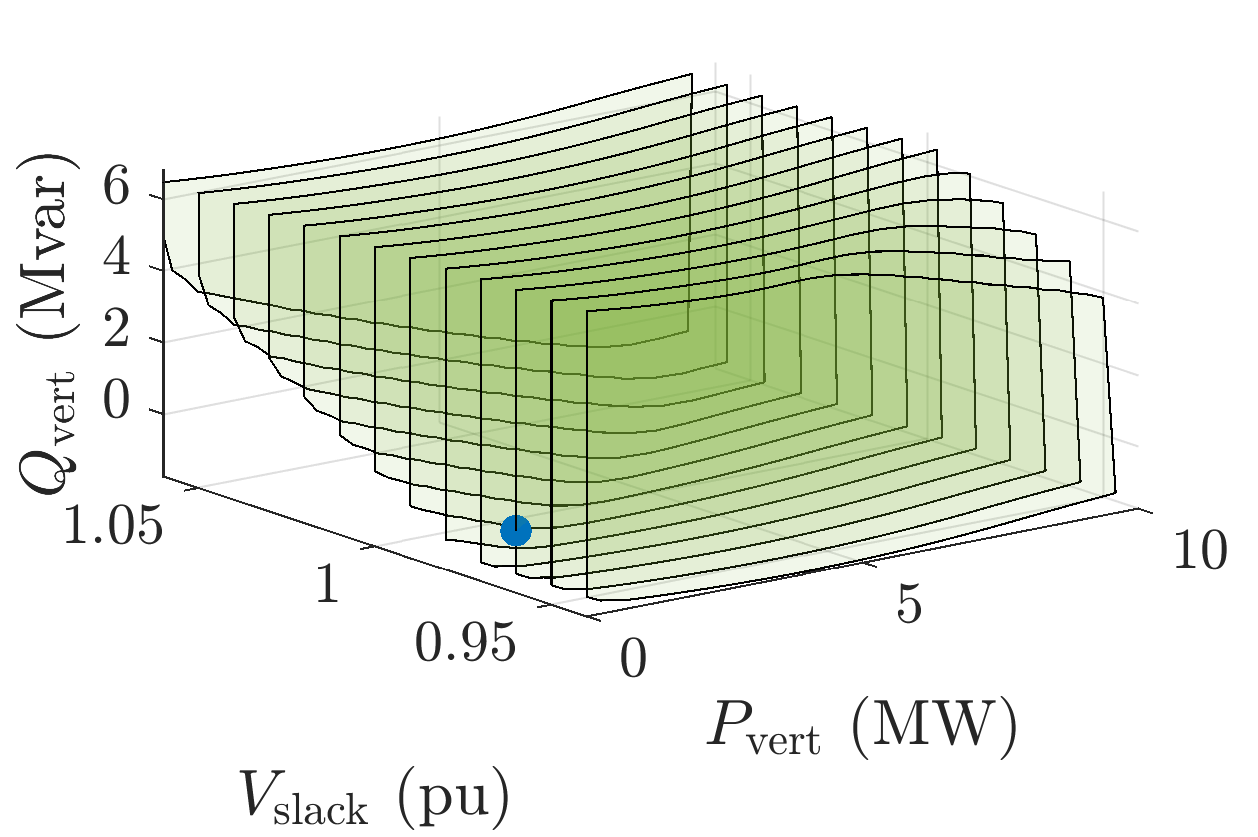}{}
    \caption{A 3D PQ(V)-FOR depicted in discrete FOR slices, from a representative adaptation of the Cigr\'e MV grid}
    \label{fig:3dFOR}
\end{figure}

The 2D representation of the FOR requires further modification considering the influence of the HV bus voltages on the underlying MV grid flexibility potentials. Significant deviations at the HV bus voltages can be caused by the flexibility activation of the aggregated FORs from the underlying MV grid levels or from the devices installed at the HV level e.g. STATCOMs, wind power plants (WPP) etc. The HV bus voltage is considered as the slack voltage for the optimal power flow (OPF) based FOR determination, as presented in \cite{majumdar2022linear}. The method is correspondingly adapted to determine the FOR for discrete slack voltages $V_{\text{slack}} \in [0.94 \ \mathrm{pu},1.06 \ \mathrm{pu}]$. An illustrative example of a 3D PQ(V)-FOR determined at the HV/MV interconnection for a representative MV grid is presented in Fig. \ref{fig:3dFOR}. The 'blue' circle represents the initial operating point. It is observed that with increased slack voltages, the potential for capacitive reactive power provision is significantly decreased at reduced values of $P_{\text{vert}}$. Increase in $P_{\text{vert}}$ signifies increased down regulation of the DER power provision, demanding increased power supply from the HV grid. This leads to increased voltage drop at the receiving end buses and correspondingly the security margin from the upper absolute bus voltage limit of 1.1 pu is improved. Accordingly, the capacitive reactive power potential is increased. Comparatively, at reduced slack voltages the inductive power potential is decreased. This is explained by the increased voltage drop at the receiving end buses, leading to a reduced security margin from the lower absolute bus voltage limit of 0.95 pu. It is noted, that the slack voltage remains constant during the power flow calculations, which is in accordance with the power flow simulations. This can be verified by inspecting the inverse Jacobian matrix sensitivities of a multi-voltage level grid. The significantly decreased sensitivities of the interconnection HV bus voltage to the MV grid flexibilities justifies a constant magnitude of the slack voltage. 

\subsection{Mixed integer formulation of the 3D PQ(V)-FOR}
\begin{figure}
    \centering
    \includegraphics[width=9cm]{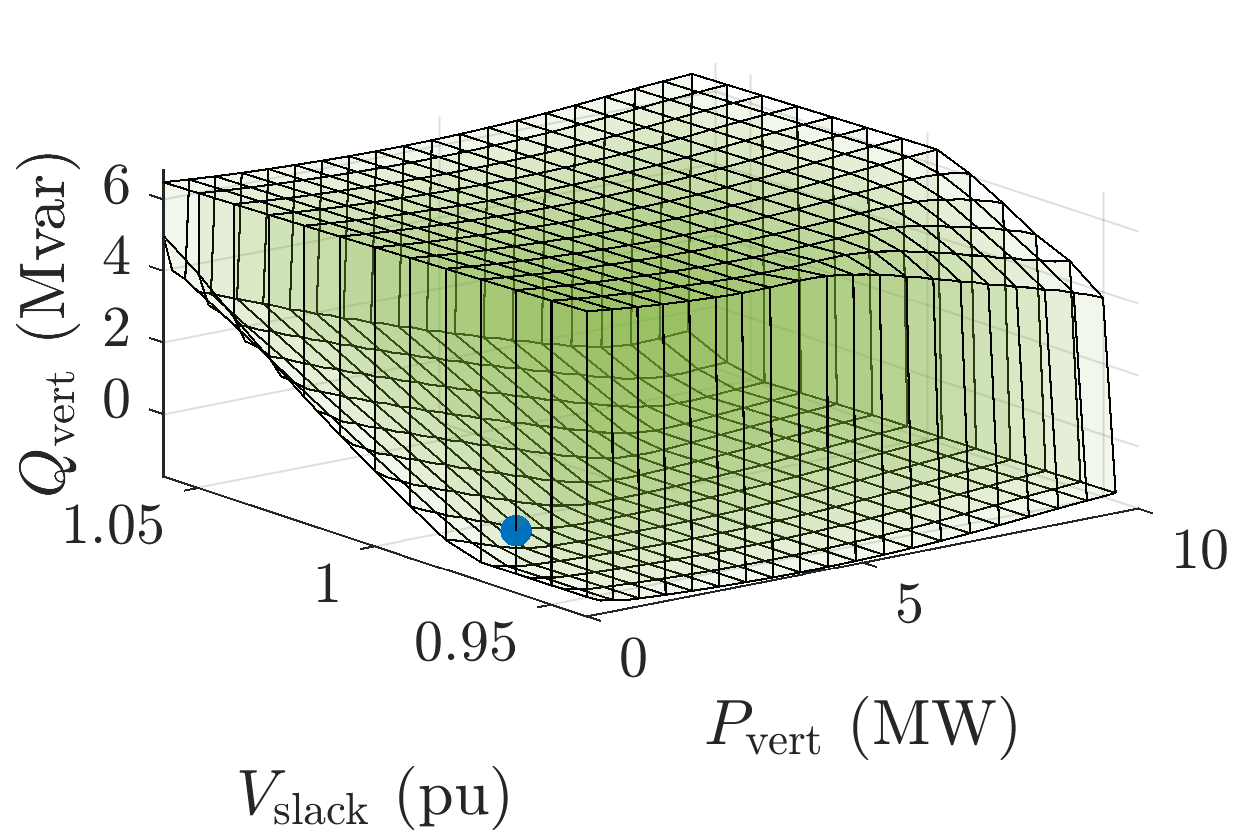}{}
    \caption{A piecewise segmentation of the 3D PQ(V)-FOR}
    \label{fig:3dFORinteger}
\end{figure}

A subsequent integration of the 3D PQ(V)-FOR in an OPF environment is required for the HV grid operational management e.g. steady state voltage control and congestion management. Therefore, the MILP formulation described in section III A.) and B.) is modified. Fig. \ref{fig:3dFORinteger} demonstrates a piecewise convex segmentation of the 3D FOR polyhedron. 
The segments are polyhedral in nature and composed of linear constraints for ease of implementation in a MILP formulation. For each segment, the upper and lower surface planar cut faces represent the dependency of the reactive power potential (Q-capability) on the active power range and the HV bus voltage considered as the slack voltage. For each bus, therefore, the total active power injection can be represented by a unique segment, selected using an integer variable. The process is described in eq. (\ref{eq:psegment1}), (\ref{eq:psegment2}). A corresponding formulation for representing the bus voltages is presented in a similar form. 
\begin{equation}
    \Delta v_i = \sum_{li=1}^{li_\text{max}}\Delta v_{li} + \sum_{li=1}^{li_\text{max}} x_{li} \cdot v_{li,\text{c,min}}
    \label{eq:vsegment1}
\end{equation}
\begin{equation}
    0 \leq \Delta v_{li} \leq x_{li} \cdot \Delta v_{li,\text{max}}
    \label{eq:vsegment2}
\end{equation}
The bus voltage segments are represented by $\Delta v_{li}$, where $li$ depicts the segment indices. The initial start point for a particular segment $li$ is represented by $v_{li,\text{c,min}}$. In the illustrative demonstration, for clarity, the segments are reduced in granularity along the $x$-axis (active power edges). In the undertaken simulations, the $x$-axis segment dimensions are further reduced by a factor of 2. This enables simplifications, as within a segment, the variance of the Q-capability with the P-capability, represented by the xz-slope is neglected. There is a negligible loss of flexibility potential, which if required can be compensated by underestimation techniques. A detailed investigation is not considered in the scope of the study. For each bus, therefore, the Q-capability is represented in relation to the slack voltage segments, depicted by the index '$li$'. Since, the Q(V)-contours are present at every active power segment interval, the index '$ki$' is used to represent the segment start points for a constant value on the $x$-axis. 
\begin{equation}
    \Delta q_{ki} \leq \sum_{li=1}^{li_\text{max}}m_{li,ki,\text{up}}\Delta v_{li,ki} + \sum_{li=1}^{li_\text{max}} q_{li,ki,\text{c,init,up}}
    \label{eq:qsegmentup_v}
\end{equation}
\begin{equation}
    \Delta q_{ki} \geq \sum_{li=1}^{li_\text{max}}m_{li,ki,\text{lo}}\Delta v_{li,ki} + \sum_{li=1}^{li_\text{max}} q_{li,ki,\text{c,init,lo}}
    \label{eq:qsegmentlow_v}
\end{equation}
The terms $m_{li,ki,\text{up}}, m_{li,ki,\text{lo}}$ represent the Q(V)-slopes along the $yz-$plane for active power segment index $ki$ along the $x-$axis. Selectivity of a unique value for '$li$' is ensured with the integer variable $x_{li}$, eq. (\ref{eq:vsegment2}). Therefore, the constraints corresponding to every active power segment is activated for the selected $li$ value. However, the Q(V)-capability is further segmented by the $x-$axis active power segments and a single segment constraint requires to be determined. Therefore, a unique decision is ensured by the integer variable $x_{ki}$ which is allocated for active power segments, eq. (\ref{eq:psegment1}), (\ref{eq:psegment2}).
\begin{equation}
    \Delta q_{ki} \leq c_\text{max}x_{ki}
    \label{eq:qc_up}
\end{equation}
\begin{equation}
    \Delta q_{ki} \geq -c_\text{max}x_{ki}
    \label{eq:qc_low}
\end{equation}
where $c_\text{max}$ is a specific constant allocated a sufficiently increased value to enable activation of the maximum/minimum PQ(V)-capability for the selected segment.

For ensuring selectivity of a single segment, the integer variables corresponding to both the intersecting active power and voltage segments is 1. Therefore, eq. (\ref{eq:integer}) is imposed for active power segments and the counterpart integer constraint for voltage segments is presented.
\begin{equation}
    \sum_{li=1}^{li_\text{max}} x_{li}=1, \quad x_{li}=[0,1]
    \label{eq:integerV}
\end{equation}
Accordingly, the deviation in bus reactive power injection/ demand is formulated by an equality constraint, where $\Delta q_{ki}$ is selected subject to a unique combination of $x_{ki}$, $x_{li}$. eq. .
\begin{equation}
    \Delta q_i = \sum_{ki=1}^{ki_\text{max}}\Delta q_{ki}
    \label{eq:qsegment_bus}
\end{equation}
Therefore, selectivity of a unique compartment is ensured by eq. (\ref{eq:qsegmentup_v})-(\ref{eq:qsegment_bus}), from which a resultant flexibility provision $f(p_i,q_i,V_\text{slack})$ can be acquired. Here, the HV bus power injections $p_i, q_i$ are derived from the aggregated 3D FOR from the underlying MV grid, depicted in Fig. \ref{fig:3dFORinteger} by the vertical active and reactive power capabilities $P_\text{vert}, Q_\text{vert}$. The equations are integrated subsequently in a MILP formulation.
\begin{equation}
\renewcommand*{\arraystretch}{1.5}
\begin{gathered}
\begin{split} \text{min} \quad & \boldsymbol{c}^\text{T}\boldsymbol{x} \mid \boldsymbol{x} = [\Delta \boldsymbol{p}^\text{T}, \Delta \boldsymbol{q}^\text{T}, \Delta \boldsymbol{\delta}^\text{T}, \Delta \boldsymbol{v}^\text{T},\boldsymbol{x}_\text{s3}^\text{T},,\boldsymbol{x}_\text{c3}^\text{T},\boldsymbol{x}_\text{int3}^\text{T}]^\text{T};\\\end{split}
\\ \begin{split}\text{s.t} \quad & \mathrm{eq. \ (\ref{eq:psegment1})-(\ref{eq:integer}),(\ref{eq:vsegment1})-(\ref{eq:qsegment_bus})}\\
& \boldsymbol{v}_{\text{min}} \leq \boldsymbol{v}_0 + \Delta \boldsymbol{v} \leq \boldsymbol{v}_{\text{max}} \\ & \boldsymbol{i}_{\text{min}} \leq \boldsymbol{i}_0 + \Delta \boldsymbol{i} \leq \boldsymbol{i}_{\text{max}}\\\end{split}
\end{gathered}\label{eq:optint3D}\end{equation}
For the MILP representation of the 3D PQ(V)-FOR, the vector term $\boldsymbol{x}_\text{s3}$ accordingly represents the set of segment variables indicated by $\Delta p_{ki}$, $\Delta q_{ki}$ and $\Delta v_{ki}$ respectively. The constant terms $p_{ki,\text{c,min}}, v_{li,\text{c,min}}$ and $q_{li,ki,\text{c,init,up}}, q_{li,ki,\text{c,init,lo}}$ are included in the vector set $\boldsymbol{x}_{c3}$. Correspondingly, the integer terms $x_{ki}, x_{li}$ are included in the vector set $\boldsymbol{x}_\text{int3}$. The optimization result provides reliable solutions in the HV grid operational management for correction of possible congestion and steady state voltage limit violation scenarios. A disadvantage of the MILP formulation is the increased computation time for determination of the optimal integer variables, and correspondingly the decisive compartments. Alternatively, as presented in the subsequent section, a convex representation is derived and integrated in a linear programming formulation. The advantages include increased computation speed for implementation in short term operational planning and reduction of optimization variables.

\section{Convexification of the 3D PQ(V)-FOR and adaptation in a linear programming formulation}
\label{sec:convexification}
\begin{figure}
    \centering
    \includegraphics[width=9cm]{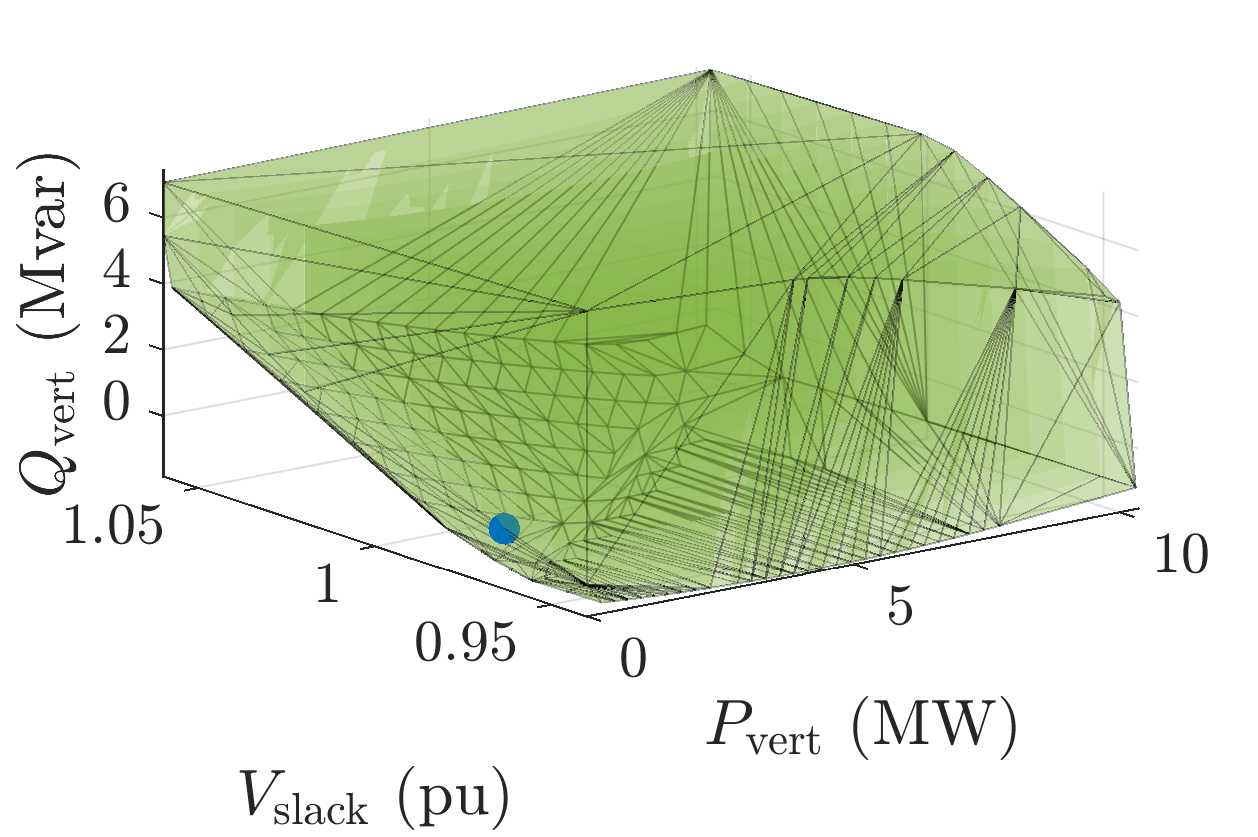}{}
    \caption{A convex representation of the 3D PQ(V)-FOR using triangulated segmentation of the convex envelope}
    \label{fig:3dFORtriangle}
\end{figure}

\begin{figure}
    \centering
    \includegraphics[width=9cm]{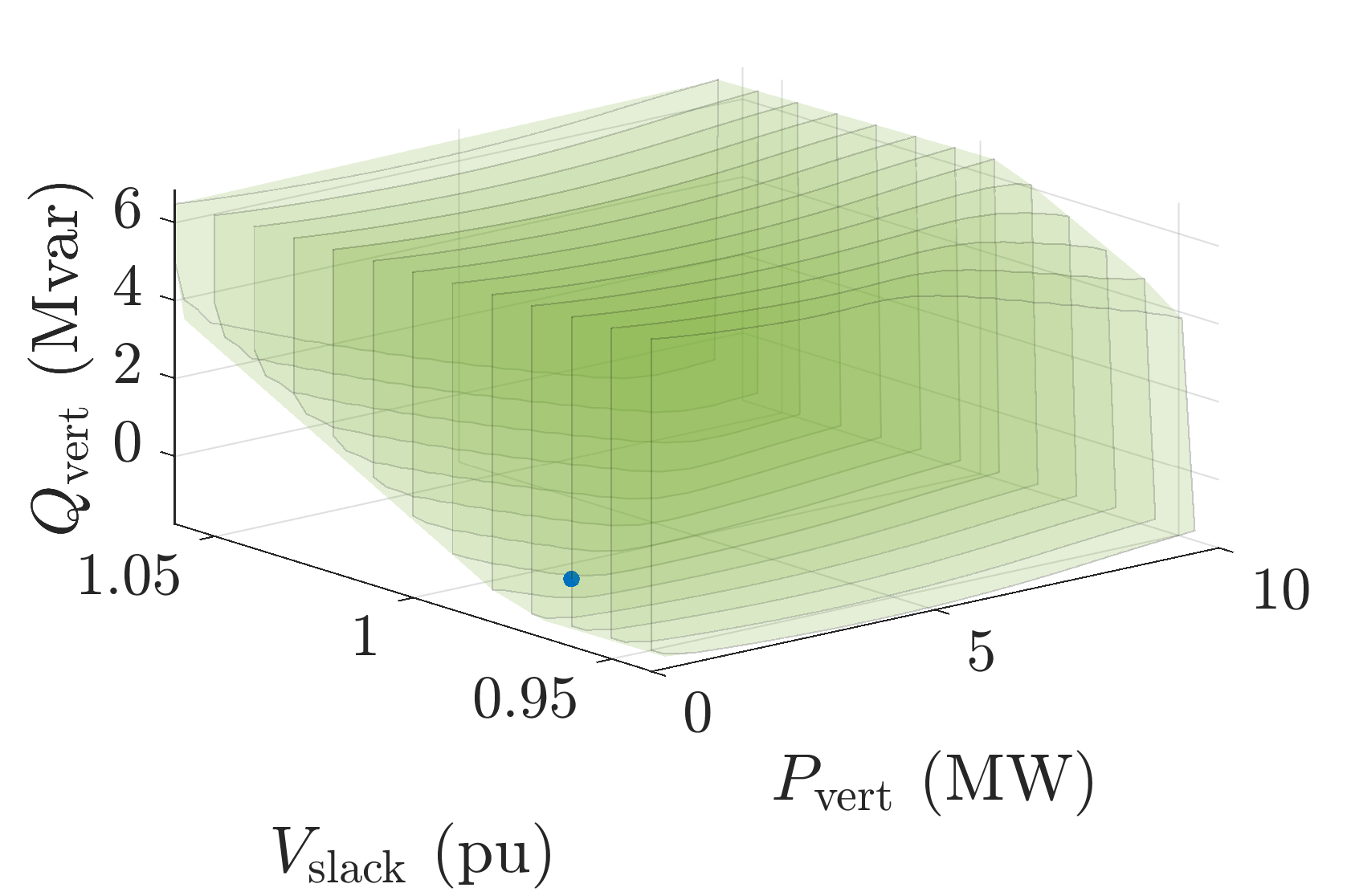}{}
    \caption{Convex hull representation of the 3D PQ(V)-FOR slices post removal of the triangle edges}
    \label{fig:3dFORenvelope}
\end{figure}

A convexified approximation of the 3D PQ(V)-FOR is presented in Fig. \ref{fig:3dFORtriangle}, \ref{fig:3dFORenvelope} \footnote{The convex hull representation is performed using the MATLAB based convhull() function, which takes as argument the incremental PQ(V) constellatory extrema points derived from the 3D PQ(V)-FOR slices}. Fig. \ref{fig:3dFORtriangle} presents a triangulated segmentation of the FOR surface area, derived from the convex hull formulation. The triangulated segments are utilized for determination of cut planes for subsequent integration in a linear programming formulation. Inspections on Fig. \ref{fig:3dFORenvelope} reveal that the convex hull presents a tight approximation of the 3D shape with negligible loss of accuracy. For example, inspection of the upper surface edge for $V_\text{slack}=$ 1.06 pu reveals an overshoot of the convex envelope as compared to the FOR slice. Similarly, an undershoot of the convex envelope from the lower surface is noted at $V_\text{slack}>$ 1 pu for initial values of $P_\text{vert}$. In such cases, a problem may arise when flexibility is acquired from the overshot or undershot sections, because the values are practically infeasible even though the deviation is marginal. Comparisons of the volumes obtained from the approximated convex hull reveal a minuscule over approximated volume of $2.45 \%$ of the actual 3D PQ(V)-FOR. The convhull() MATLAB function provides a parameter for a direct calculation of the convex hull volume. For determining the volume of the 3D PQ(V)-FOR, the volume of the polyhedral segments as depicted in Fig. \ref{fig:3dFORinteger} are calculated using the convhull() function and cumulatively summed up. For increased accuracy, a corresponding conservative convex hull representation with an assured security margin is required. A complete survey on the determination of such conservative approximation strategies is beyond the scope of this paper.  

\subsection{Linear programming formulation of the convexified 3D PQ(V)-FOR using cut plane representation of the surface area}
\begin{figure}
    \centering
    \includegraphics[width=8cm]{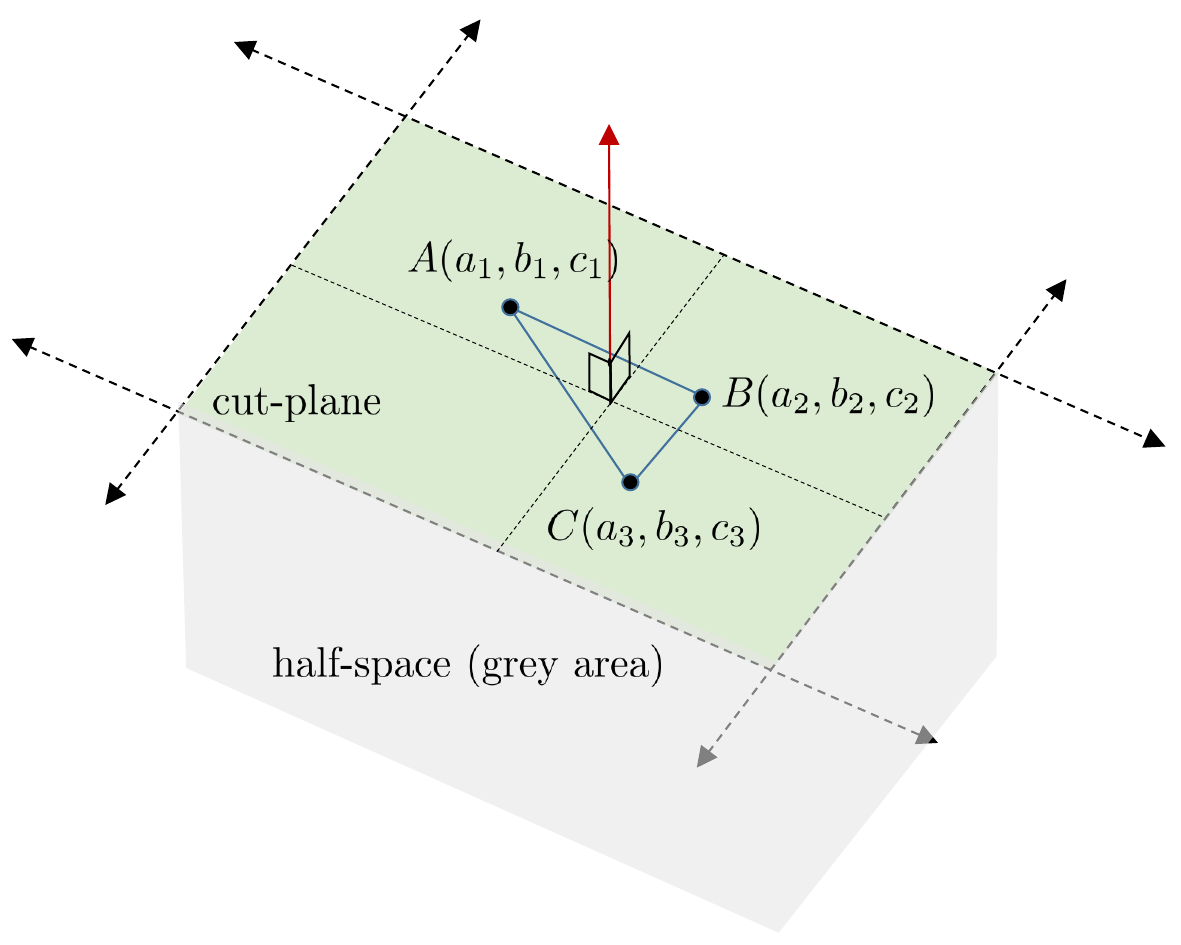}{}
    \caption{Schematic representation of a half-space and normal obtained from 3 points on a cut-plane}
    \label{fig:halfspace}
\end{figure}

Cut planes can be derived from the triangulated segments on the surface area and corresponding half spaces can be formulated. The convex hull representation presents an intersection of the permissible half spaces. The vertices from the triangle segments are used to describe plane equations. A schematic representation is presented in Fig. \ref{fig:halfspace}. A plane section is depicted with an embedded triangle segment. The normal to the plane and the underlying half-space is derived from the planar equations obtained from the non colinear vertices of the triangle. The black bidirectional arrows depict the extension of the plane across all directions. The cut-plane splits a 3D space into two half-spaces.
The equation of the normal vector $n_\text{p}$ to the cut-plane is obtained from the cross product of two coplanar vectors.
\begin{equation}
    \vec{n_\text{p}}=\vec{AB} \times \vec{BC} = (n_\text{p1},n_\text{p2},n_\text{p3})
\end{equation}
where $(n_\text{p1},n_\text{p2},n_\text{p3})$ represent the $x,y,z$-axes coefficients of the normal vector $\vec{n_\text{p}}$. Therefore, the plane equation can be represented.
\begin{equation}
    n_\text{p1}x+n_\text{p2}y+n_\text{p3}z=d
    \label{eq:plane}
\end{equation}
The constant term $d$ is obtained by inserting the $x,y,z$-coordinates of a point e.g. $A (a_1, b_1, c_1)$ lying on the plane into the plane equation, eq. (\ref{eq:plane}).
\begin{equation}
    d=n_\text{p1}a_1+n_\text{p2}b_1+n_\text{p3}c_1
\end{equation}
The corresponding underlying half-space can be determined by an affine inequality.
\begin{equation}
    n_\text{p1}x+n_\text{p2}y+n_\text{p3}z \leq d
\end{equation}
Therefore, the half-spaces for all triangular segments are derived and integrated in a linear inequality formulation. For bus index $i$, the corresponding half-space inequalities in a linear programming formulation are presented.
\begin{equation}
    n_{\text{p1},ti}\Delta p_i+n_{\text{p2},ti}\Delta q_i+n_{\text{p3},ti}\Delta v_i \leq d_{ti} \ \forall \ ti \in t
    \label{eq:halfspace_tri}
\end{equation}
The index '$ti$' indicates the triangular segments for bus '$i$' and '$t$' represents the total number of triangular segments for the convex envelope. Therefore, the 3D PQ(V)-FOR are not described by affine half-spaces. The number of optimization variables representing the grid state vectors $\boldsymbol{x}=[\Delta \boldsymbol{p}^\text{T}, \Delta \boldsymbol{q}^\text{T}, \Delta \boldsymbol{\delta}^\text{T}, \Delta \boldsymbol{v}^\text{T}]^\text{T}$ remains unchanged. However, numerous constraints from the half-space inequalities for each triangular segment $ti$ at each bus $i$ are introduced.
\begin{equation}
\renewcommand*{\arraystretch}{1.5}
\begin{gathered}
\begin{split} \text{min} \quad & \boldsymbol{c}^\text{T}\boldsymbol{x} \mid \boldsymbol{x} = [\Delta \boldsymbol{p}^\text{T}, \Delta \boldsymbol{q}^\text{T}, \Delta \boldsymbol{\delta}^\text{T}, \Delta \boldsymbol{v}^\text{T}]^\text{T}_{m,1};\\ & \boldsymbol{c}=[c_k]^\text{T}_{m,1}, k = [1,m] \cap \mathbb{Z}; m = 4n\end{split}
\\ \begin{split}\text{s.t} \quad & n_{\text{p1},ti}\Delta p_i+n_{\text{p2},ti}\Delta q_i+n_{\text{p3},ti}\Delta v_i \leq d_{ti} \ \forall i \in n\\
& \boldsymbol{v}_{\text{min}} \leq \boldsymbol{v}_0 + \Delta \boldsymbol{v} \leq \boldsymbol{v}_{\text{max}} \\ & \boldsymbol{i}_{\text{min}} \leq \boldsymbol{i}_0 + \Delta \boldsymbol{i} \leq \boldsymbol{i}_{\text{max}}\\\end{split}
\end{gathered}\label{eq:optint3D}\end{equation}

\section{Results: Application of the 3D PQ(V)-FORs for the HV grid operational management}
\begin{figure}
    \centering
    \includegraphics[width=9cm]{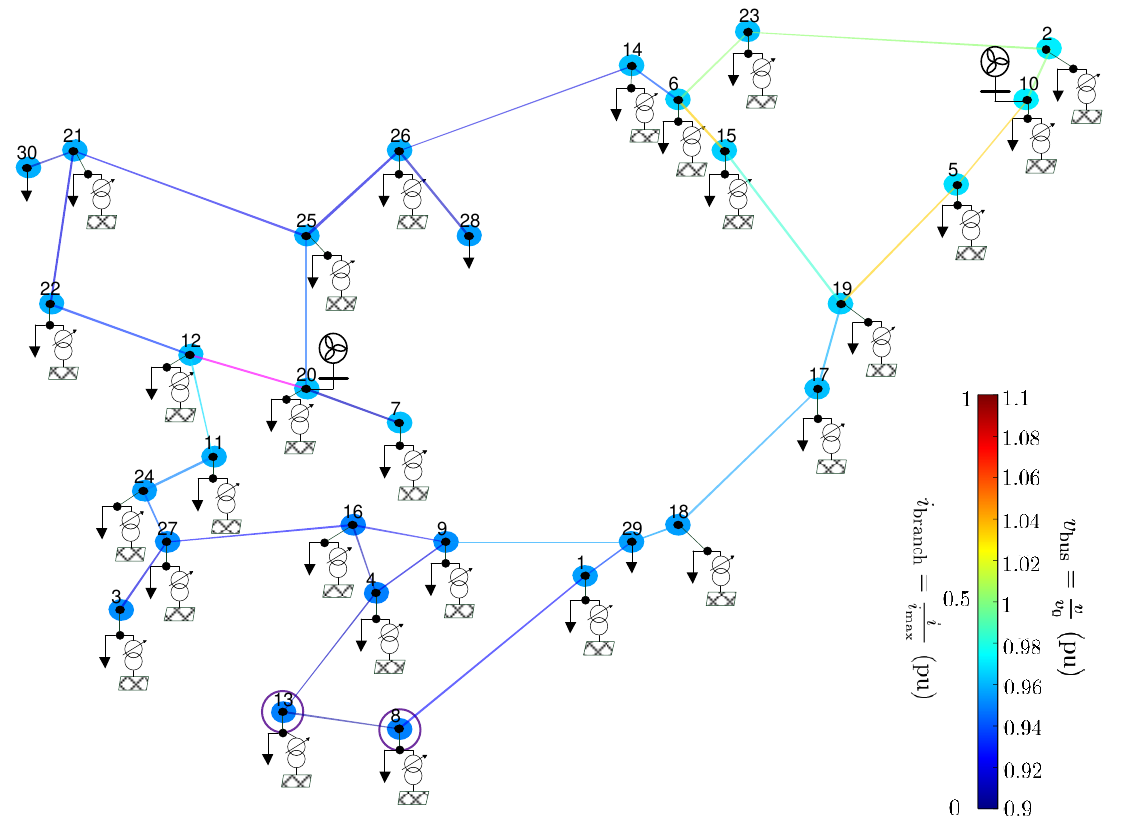}{}
    \caption{Grid topology demonstrating initial scenario for a congestion and bus voltage limit violation}
    \label{fig:gridscenario}
\end{figure}

The application of the 3D PQ(V)-FORs are tested on a EHV/HV grid topology as demonstrated in Fig. \ref{fig:gridscenario}. The grid is derived from a multi-voltage level power system format with equipped dataset, presented in \cite{sarstedt2019modelling}\cite{ifes-eevdataset}. The original grid consists of 30 HV buses and 3 EHV buses (EHV1-EHV3). However, for mathematical optimization and power flow calculations, the EHV buses are connected to a common EHV bus (EHV4) considered as a slack, demonstrated in \cite{majumdar2023reliability}. The initial grid scenario is depicted, where the bus voltage and branch current profiles are presented using a color spectrum. The 'magenta' colored branch indicates a congested line. The bus voltage limit breaches are indicated by hollowed circles for bus id: 8, 13. HV wind parks and aggregated loads at the HV buses are accordingly represented. The underlying MV grid levels are illustrated by external grid connected by HV/MV transformers. The application of the 3D PQ(V)-FOR for HV grid operational management is demonstrated in the following subsections for the MILP based formulation and the convexification approach. The optimized solutions are compared to the initial grid states and the corrective action is demonstrated in the subsequent sections. Representative 3D PQ(V)-FORs are correspondingly presented to exhibit deviation of the operating point from the initial within the bounds of the FORs.
\subsection{Optimized voltage and current profiles based on the MILP and convexificated formulations of the 3D PQ(V)-FOR}
\begin{figure}
    \centering
    \includegraphics[width=9cm]{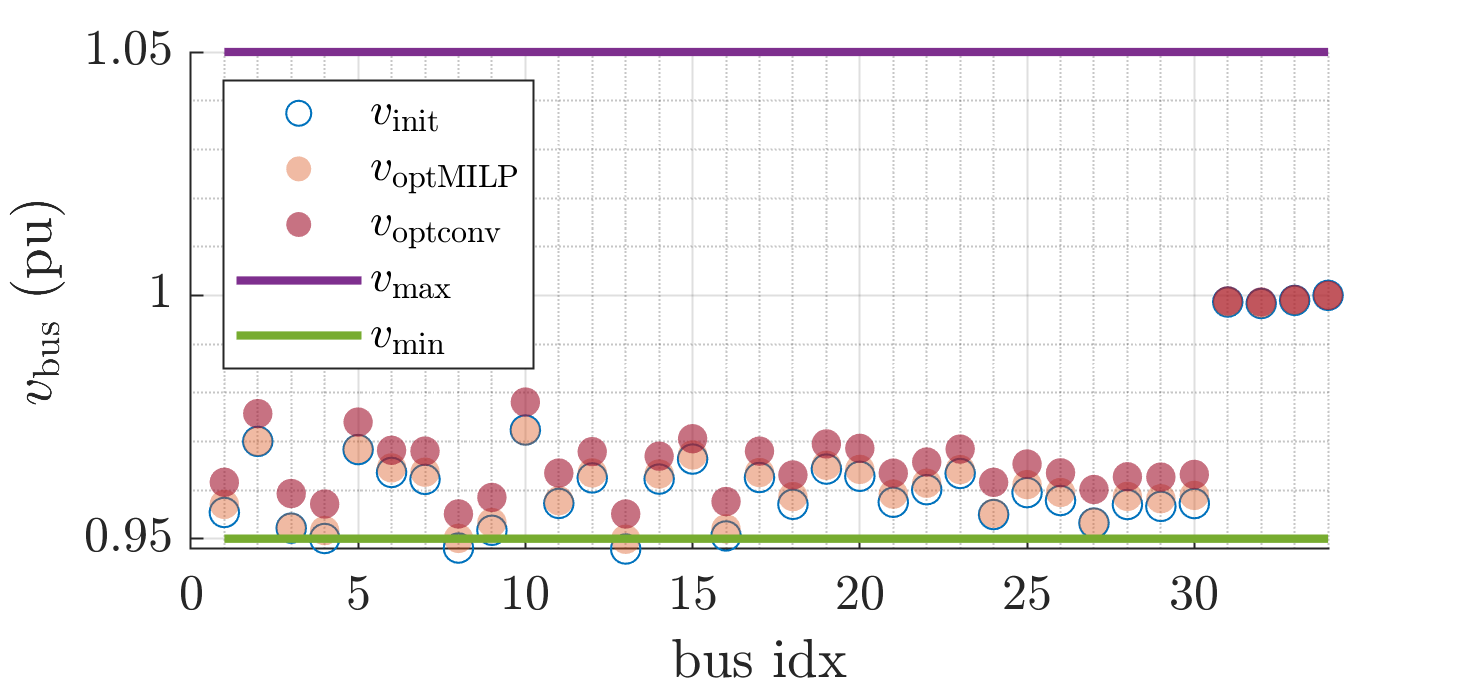}{}
    \caption{Bus voltage magnitude comparison between the initial scenario and the optimized solutions}
    \label{fig:volcompare}
\end{figure}

\begin{figure}
    \centering
    \includegraphics[width=9cm]{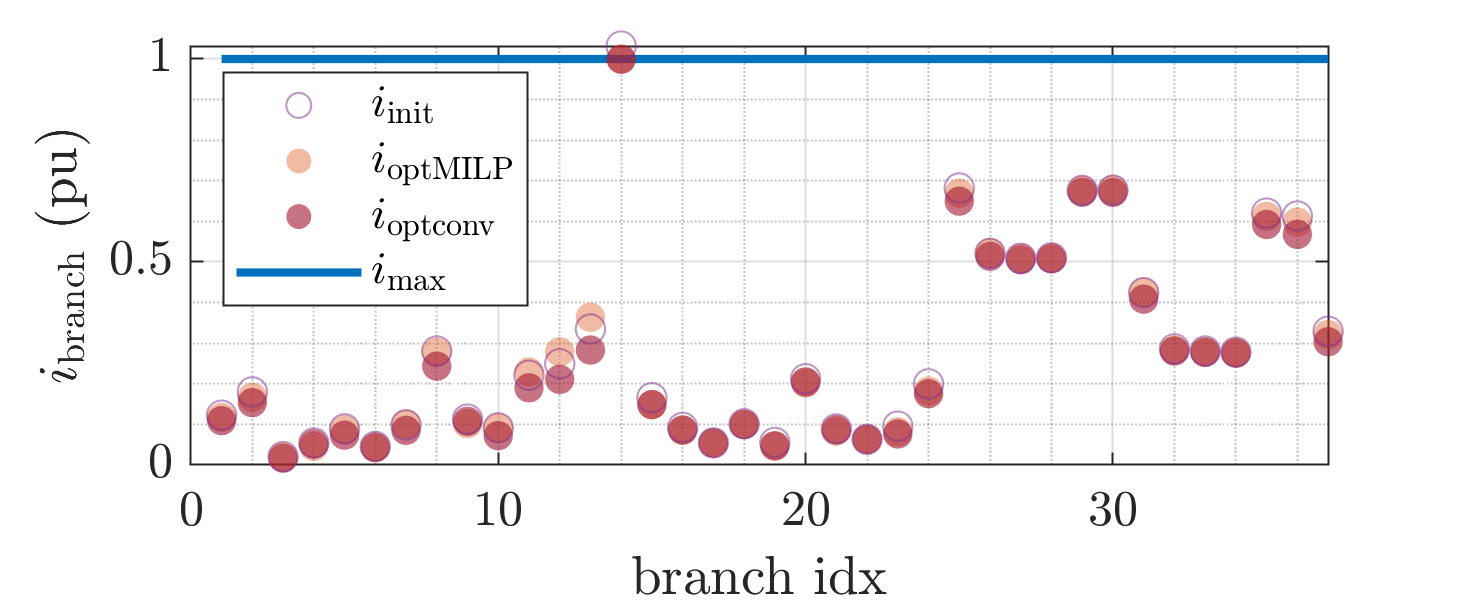}{}
    \caption{Branch current magnitude comparison between the initial scenario and the optimized solutions}
    \label{fig:currcompare}
\end{figure}

The initial state of the bus voltages $v_\mathrm{init}$ and the optimized results ($v_\mathrm{optMILP}$, $v_\mathrm{optconv}$) are presented in Fig. \ref{fig:volcompare}. An inspection of bus id: 8, 13 reveals the corrected voltage profiles within the absolute maximum ($v_\mathrm{max}$) and minimum ($v_\mathrm{min}$) voltage constraints. The subscripts in the legend 'optMILP' and 'optconv' refer to the optimal solutions obtained from the MILP and convexification based approaches respectively. The convexification based solution indicates an increased rise in the voltage profile, signifying increased capacitive reactive power demand at the buses. Accordingly, the branch current profiles normalized with respect to the maximum thermal branch current constraints are presented in Fig. \ref{fig:currcompare}.  Observations at the branch id: 14 reveals a correction of the current profile using both the MILP and convexification based approaches. Inspite of different solutions being procured, the reliability of HV grid operational management is achieved in both cases. The results are obtained by inserting the optimal solutions of the bus active and reactive power changes during the power flow calculations. In case of discrepancies, where the optimal solutions fail to satisfy the grid constraints during power flow calculations, subsequent iterations are recommended. The corresponding section illustrates the flexibility procurement from the 3D PQ(V)-FORs using the MILP and convexification based approaches.  

\subsection{Flexibility procurement from the 3D PQ(V)-FORs based on the MILP and convexification based approaches}
\begin{figure}
	\centering
	\begin{tabular}{cc}
		\subfloat[bus id=7]{\includegraphics[width=0.22\textwidth]{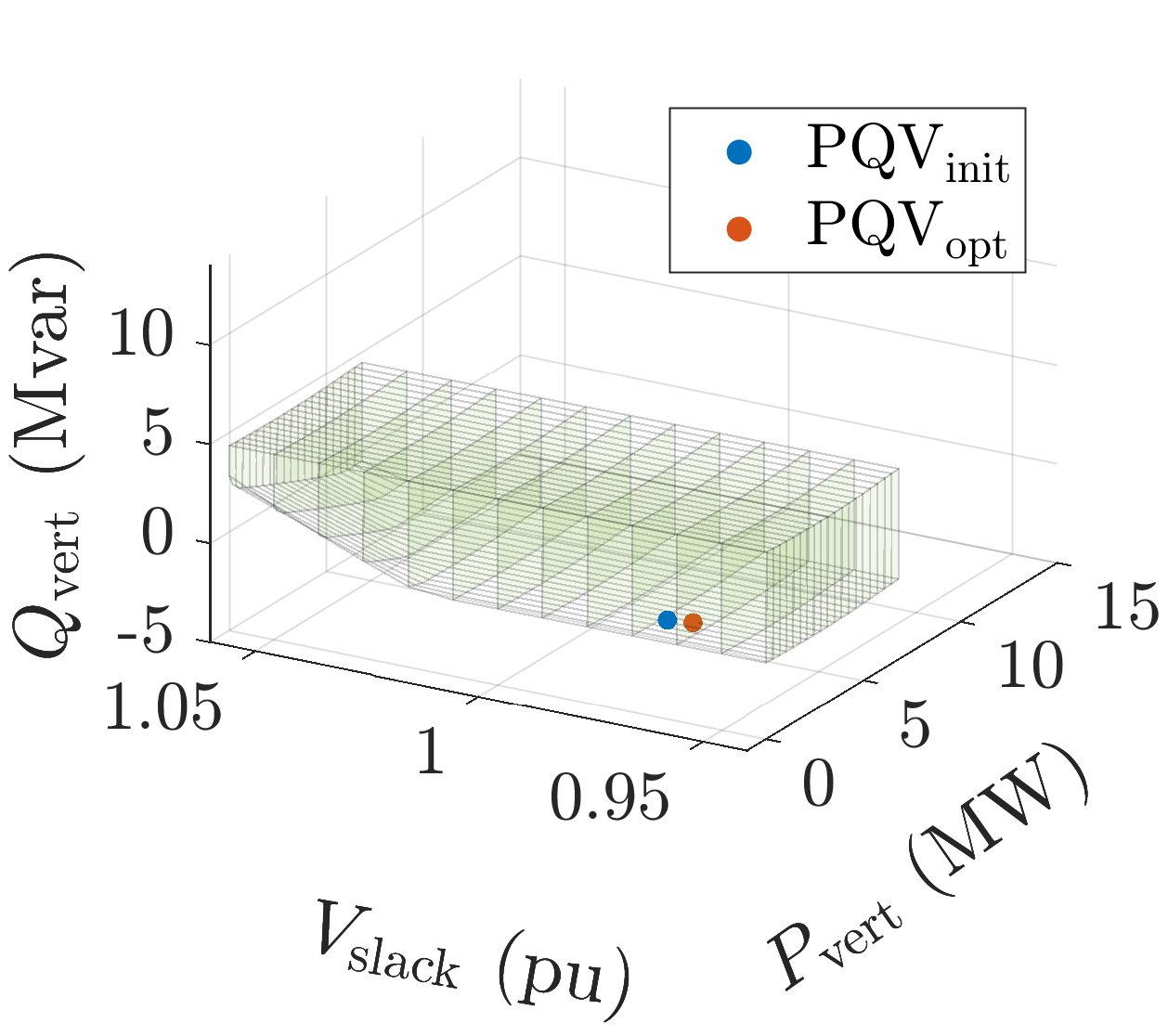}}
		&
		\subfloat[bus id=20]{\includegraphics[width=0.22\textwidth]{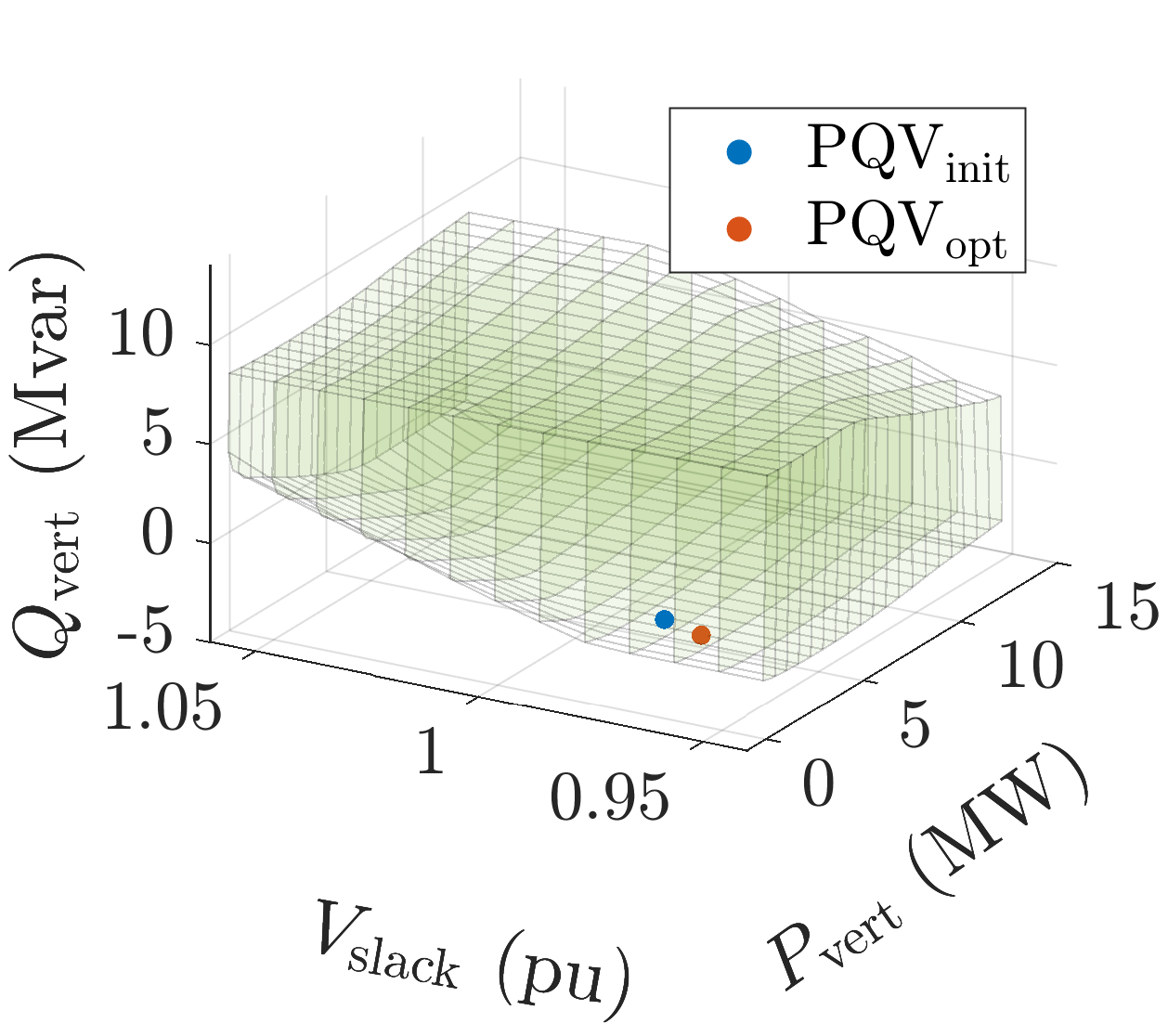}}
	\end{tabular}
	\caption{Optimal flexibility acquisition from the 3D PQ(V)-FOR using the MILP based method}
	\label{fig:integerFOR}
\end{figure}

\begin{figure}
	\centering
	\begin{tabular}{cc}
		\subfloat[bus id=7]{\includegraphics[width=0.22\textwidth]{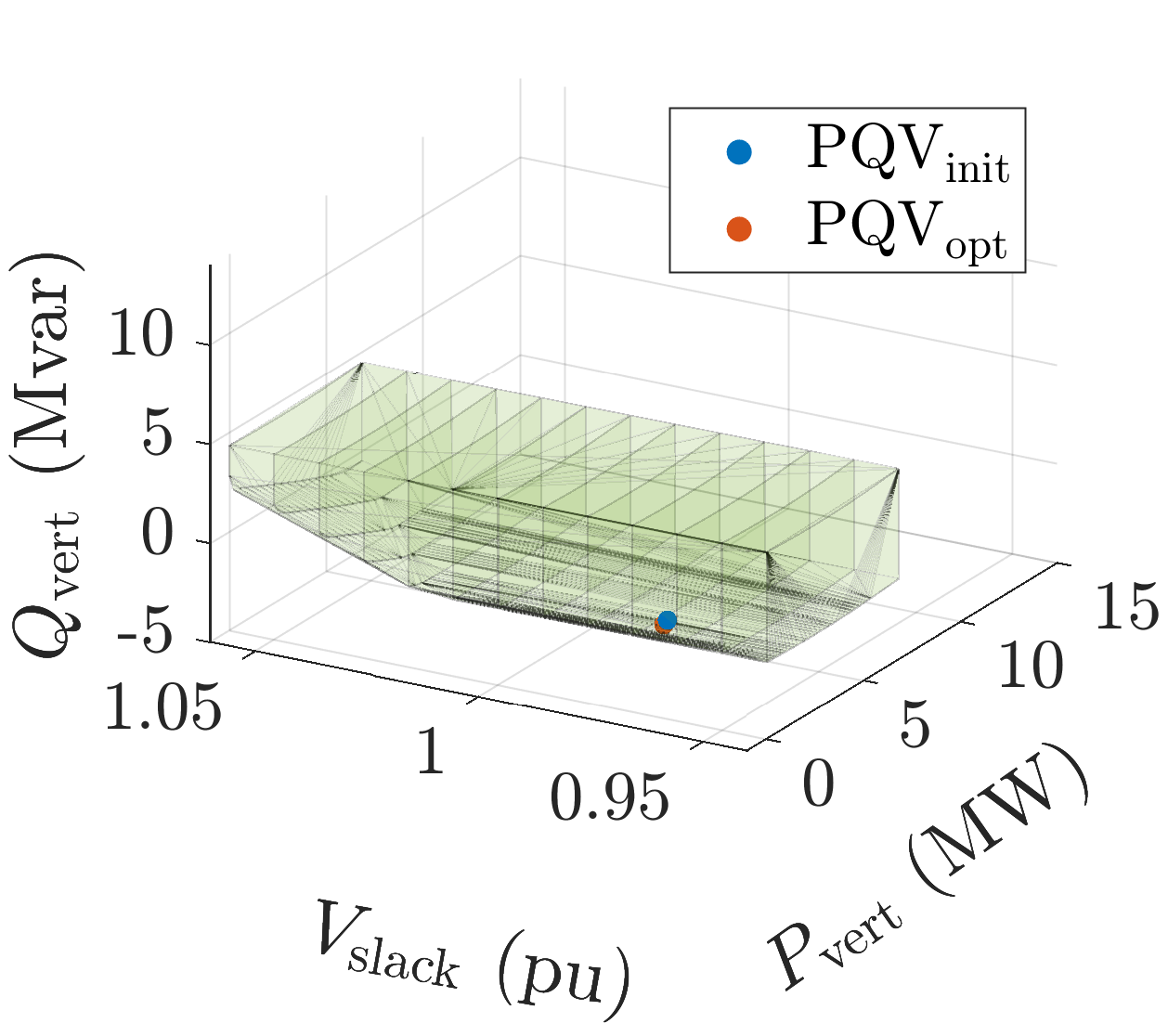}}
		&
		\subfloat[bus id=20]{\includegraphics[width=0.22\textwidth]{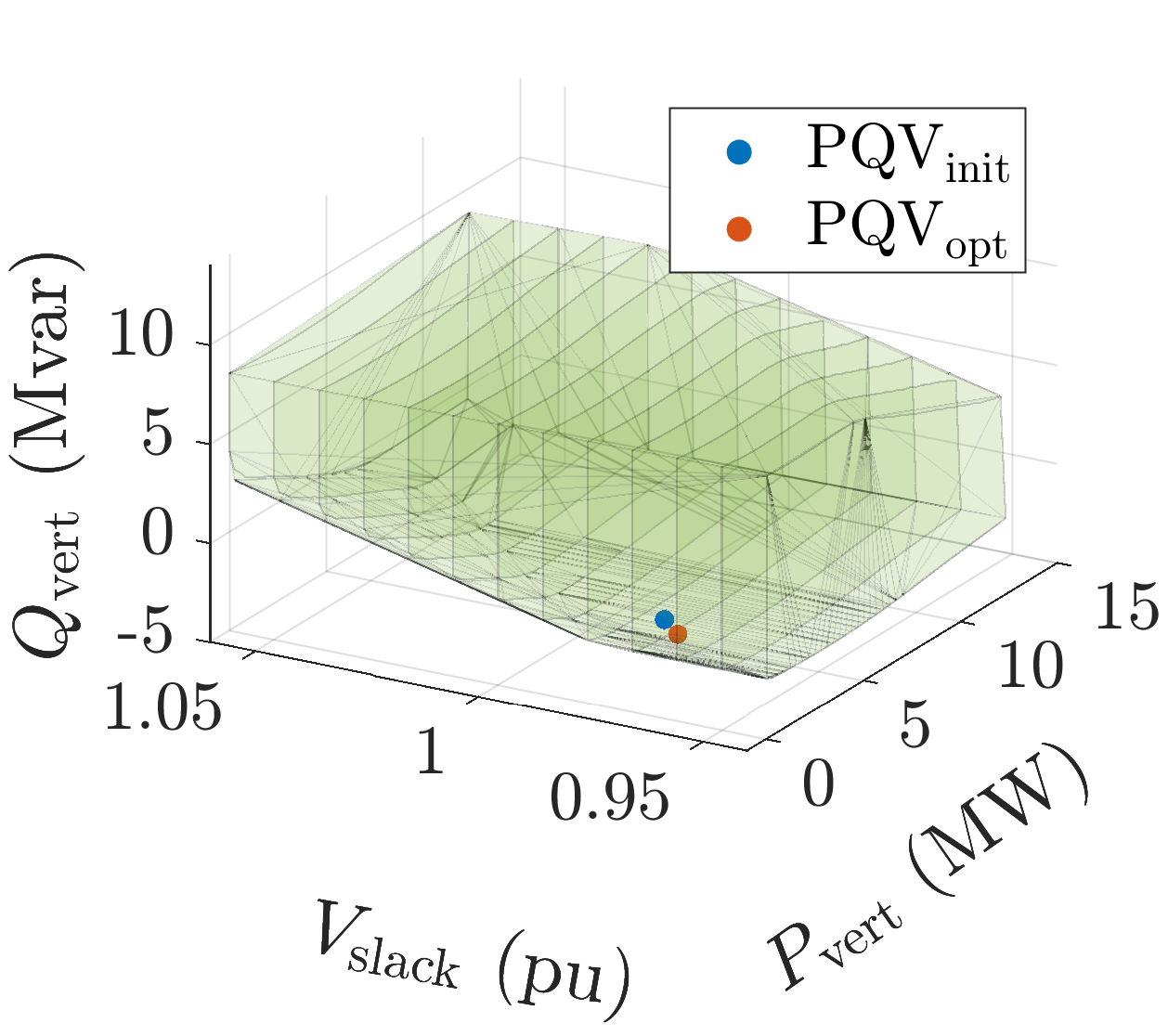}}
	\end{tabular}
	\caption{Optimal flexibility acquisition from the 3D PQ(V)-FOR using the convexification based linear programming}
	\label{fig:convFOR}
\end{figure}

\begin{figure}
    \centering
    \includegraphics[width=9.2cm]{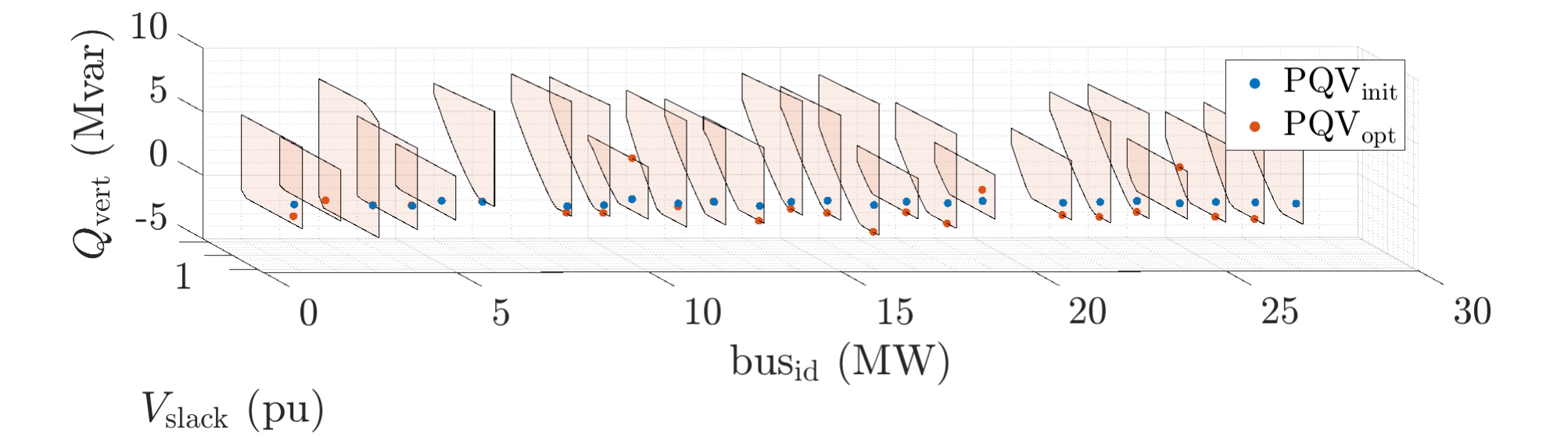}{}
    \caption{Optimal reactive power flexibility provision with vertical demand of $\Delta P_\mathrm{vert}$=0 MW, using MILP based method}
    \label{fig:QVintegerFOR}
\end{figure}

\begin{figure}
    \centering
    \includegraphics[width=9.2cm]{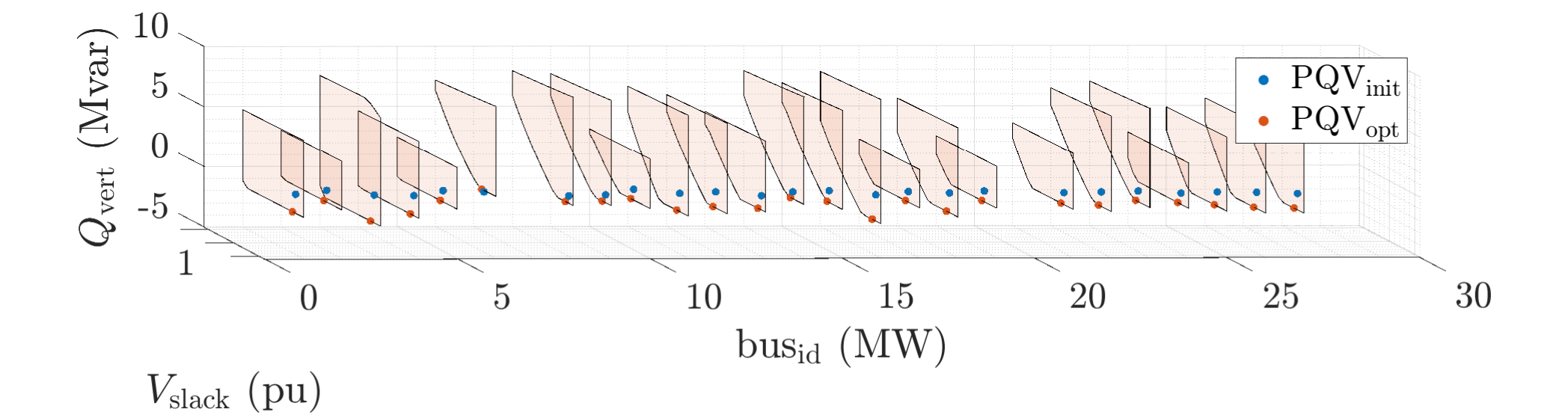}{}
    \caption{Optimal reactive power flexibility provision with vertical demand of $\Delta P_\mathrm{vert}$=0 MW, using convexification method}
    \label{fig:QVconvFOR}
\end{figure}

The flexibility procurement using both active and reactive power are presented in Fig. \ref{fig:integerFOR}, \ref{fig:convFOR} corresponding to the MILP and convexification based approaches respectively. Bus id: 7, 20 are in close proximity to the congested branch id: 14 as evident from Fig. \ref{fig:gridscenario}. Therefore, PQ(V)-flexibility procurement from the underlying grid 3D PQ(V)-FORs at the buses are required to correct the constraint violation. Observations further reveal that the adapted operating points lie within the bounds prescribed by the polyhedral 3D PQ(V)-FORs in both cases. Furthermore, Q(V)-flexibility procurement from the other constituent buses of the grid is presented in Fig. \ref{fig:QVintegerFOR}, \ref{fig:QVconvFOR} respectively for the MILP and the convexification based solutions. Since, reactive power procurement is allocated zero costs in the objective function, the optimizer prioritizes procurement of Q(V)-flexibility from the 3D PQ(V)-FORs. The corresponding active power set point at bus $i$ (excluding 7, 20) is unchanged ($\Delta P_{\mathrm{vert},i}=0$). Therefore, the two dimensional Q(V)-FOR slices ($\Delta P_{\mathrm{vert},i}=0$) are presented for both cases, with the corresponding initial and optimized operating points. Bus ids: 7, 20 are excluded as additional active power procurement is demonstrated using the 3D PQ(V)-FORs, presented in Fig. \ref{fig:integerFOR}, \ref{fig:convFOR}. The general trend is towards the production of capacitive reactive power, as apparent from the optimized operating point. The adapted solution converges towards the lower edges of the respective FOR slices signifying a capacitive reactive power contribution. This is attributed to the congestion management in addition to the undervoltage correction at bus id: 8, 13.

A comparison of computation times is presented in Table \ref{tab:computation times}. Evidence depicts fast computations times of the convexification approach by a factor of 120 as compared to the MILP based segmentation. Therefore, the linear programming based strategy can be used for real time simulations, to quantify and procure flexibilities in short term intervals. However, the accuracy of the method also ensures validity in intraday/ day-ahead operational planning.
\begin{table}[]
\caption{Computation times for 3D PQ(V)-FOR determination strategies}
\begin{tabular}{lll}
\rowcolor[HTML]{A7F3A6} 
Methods           & Segmentation (MILP)       & Convexification (LP)        \\
Computation times & \multicolumn{1}{c}{1203 s} & \multicolumn{1}{c}{9.5 s}
\end{tabular}
\label{tab:computation times}
\end{table}

\subsection{Flexibility procurement from the 3D PQ(V)-FORs under consideration of increased grid state deviations}
The flexibility procurement is assured for small scale grid state (e.g. bus voltages) deviations. However, the response to increased deviations requires consideration as a linear programming based approach loses accuracy with increased deviations from the initial operating point. Furthermore, shunt switching for example, in the proximity of the affected buses can improve the voltage profile of the grid and correspondingly relieve congestion at an affected branch. Accordingly active power curtailment from the aggregated PQ(V) flexibilities from the underlying distribution grids can be minimized. Therefore, economic efficiency is prioritised, since, active power curtailment signifies loss of opportunity to engage cheaper DERs from the distribution grid levels. 

\begin{figure*}[!t]
\centering
\subfloat[]{\includegraphics[width=1.25in]{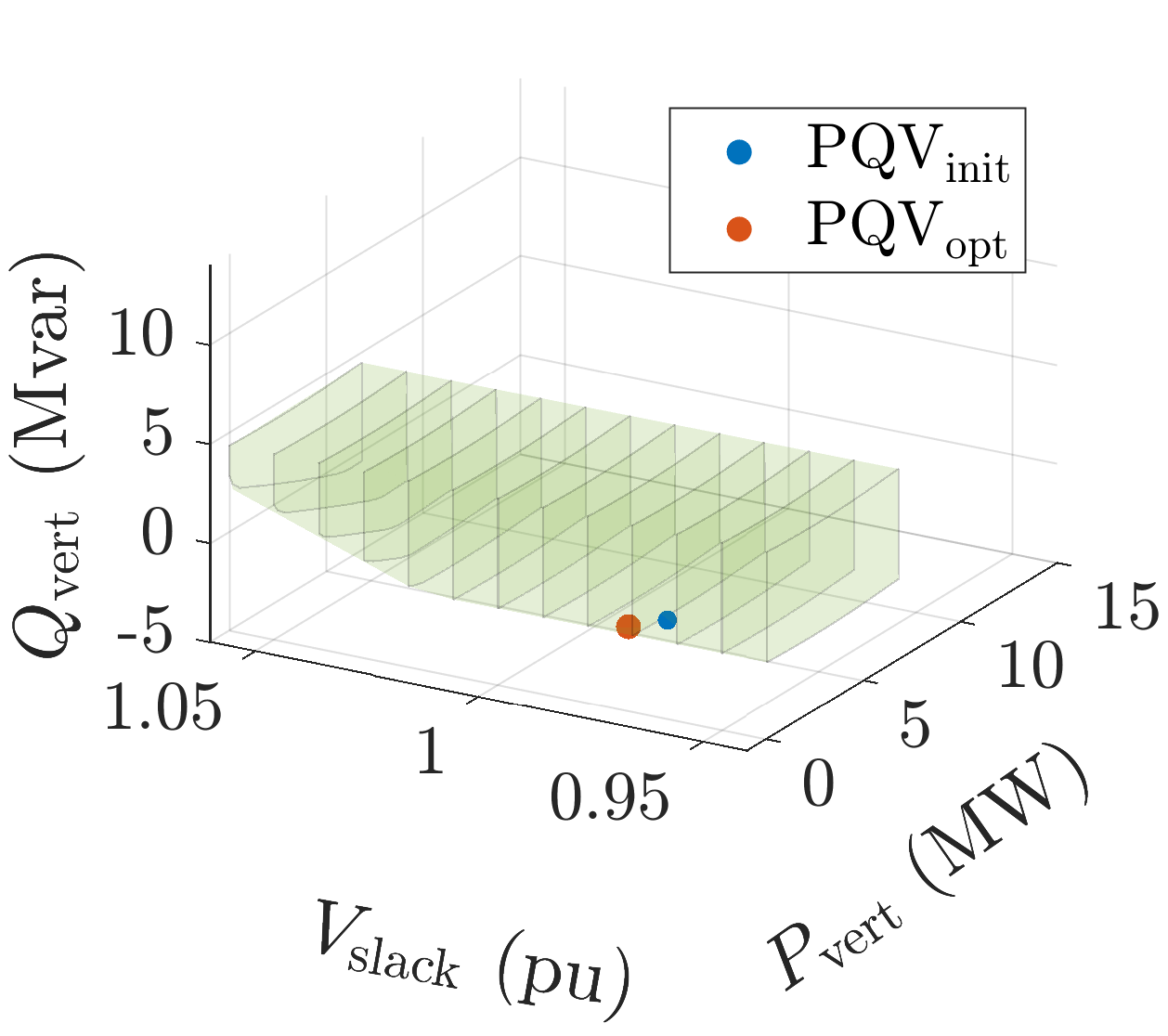}%
\label{fig_first_case}}
\hfil
\subfloat[]{\includegraphics[width=1.25in]{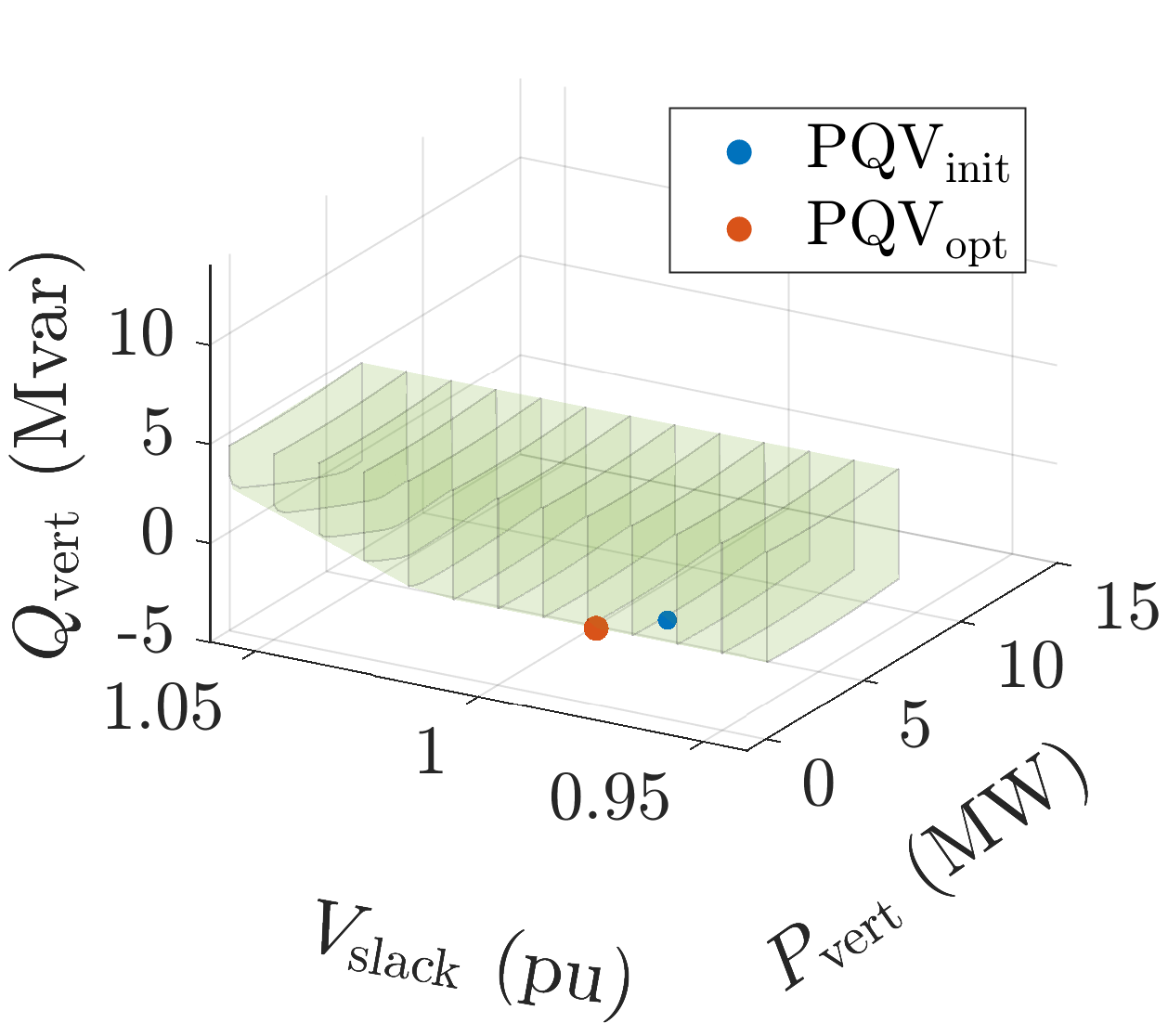}%
\label{fig_second_case}}
\hfil
\subfloat[]{\includegraphics[width=1.25in]{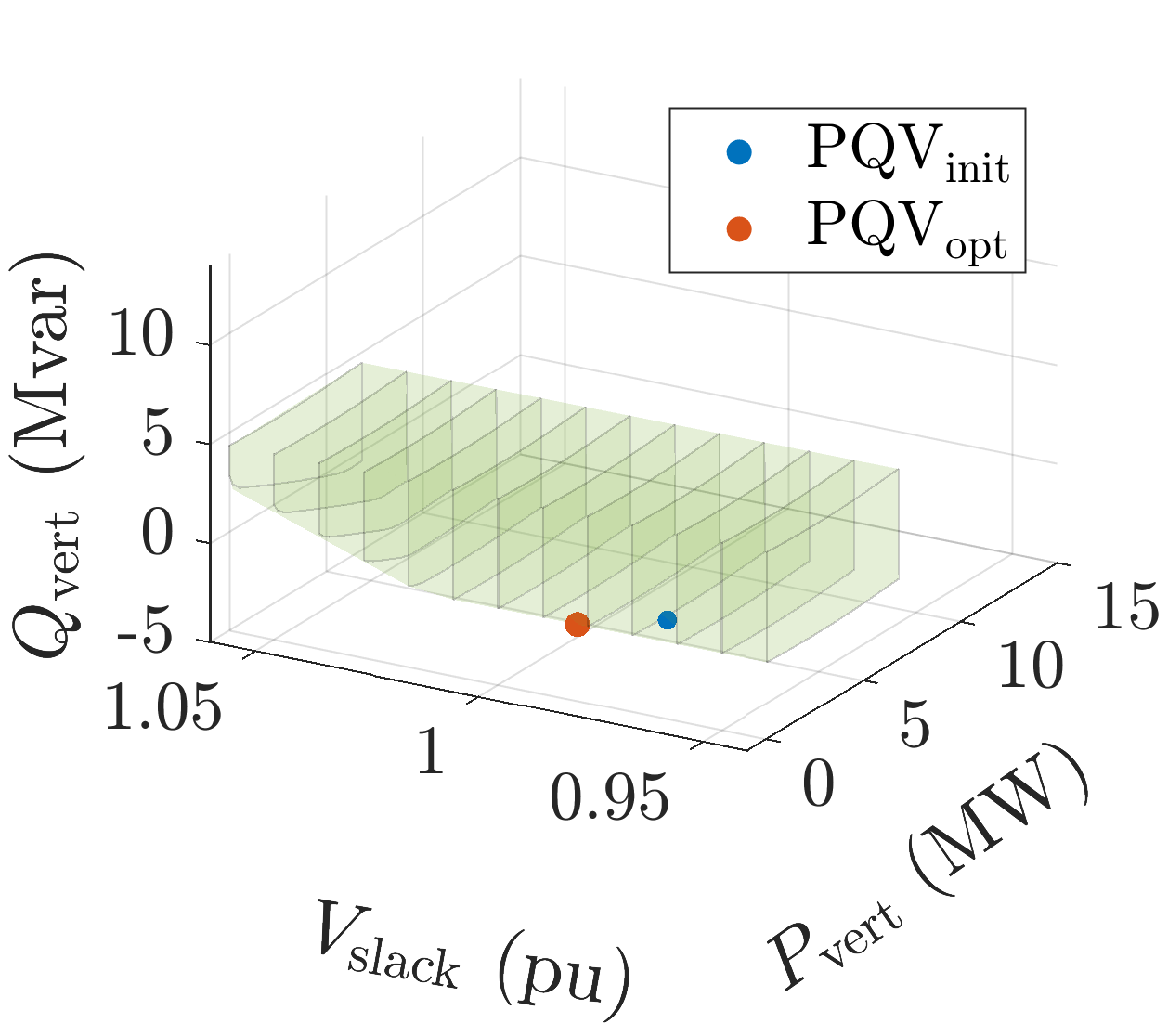}%
\label{fig_second_case}}
\hfil
\subfloat[]{\includegraphics[width=1.25in]{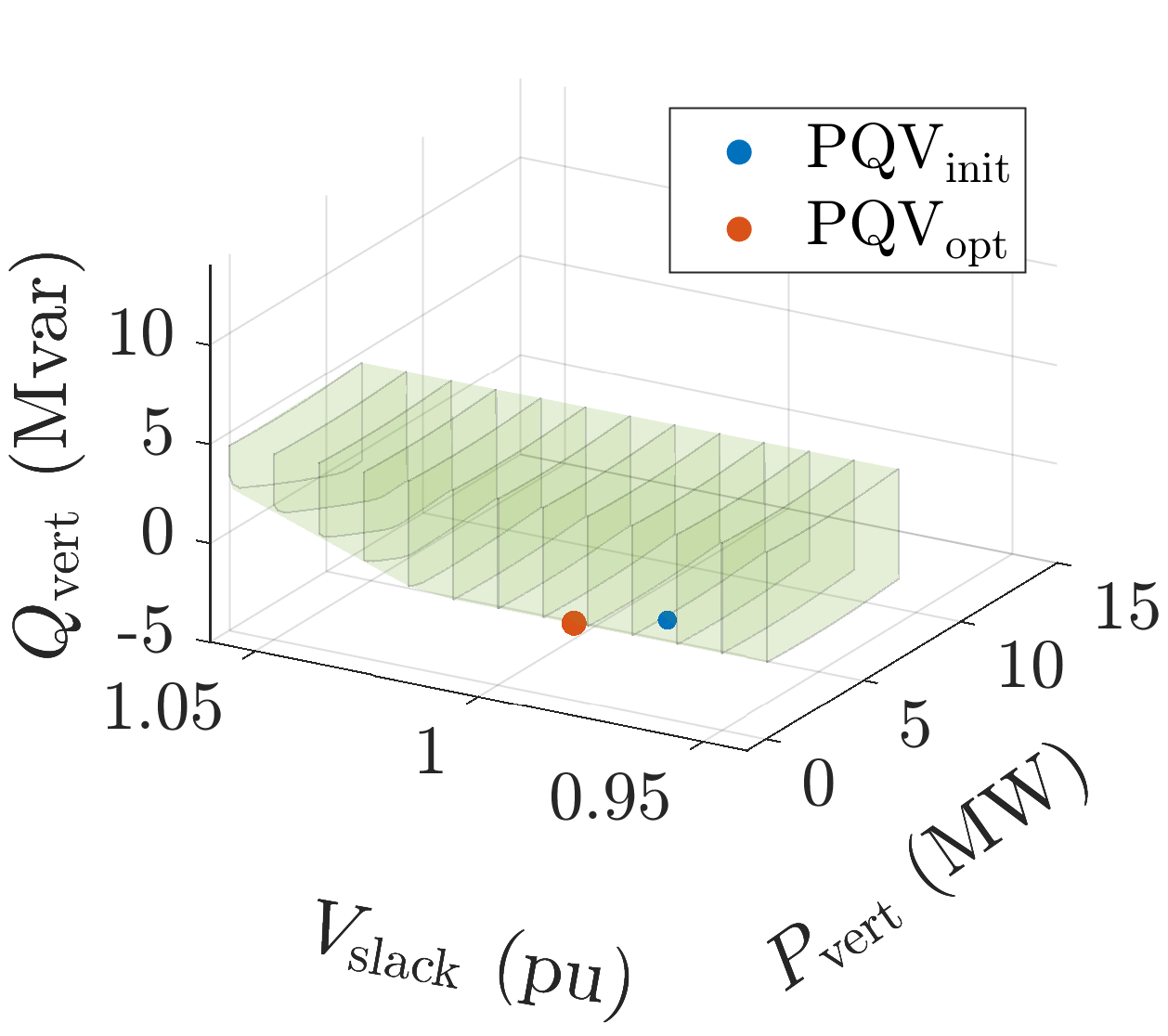}%
\label{fig_second_case}}
\caption{Deviation of the PQ(V) operating point at bus id=7 in response to increased deviations due to reactive power injection: (a) $Q_\mathrm{bus \ id,29}=-50 \ \mathrm{Mvar}$ (b) $Q_\mathrm{bus \ id,29}=-100 \ \mathrm{Mvar}$ (c) $Q_\mathrm{bus \ id,29}=-150 \ \mathrm{Mvar}$ (d) $Q_\mathrm{bus \ id,29}=-200 \ \mathrm{Mvar}$.}
\label{fig:fig_Qbigbus7}
\end{figure*}

\begin{figure*}[!t]
\centering
\subfloat[]{\includegraphics[width=1.25in]{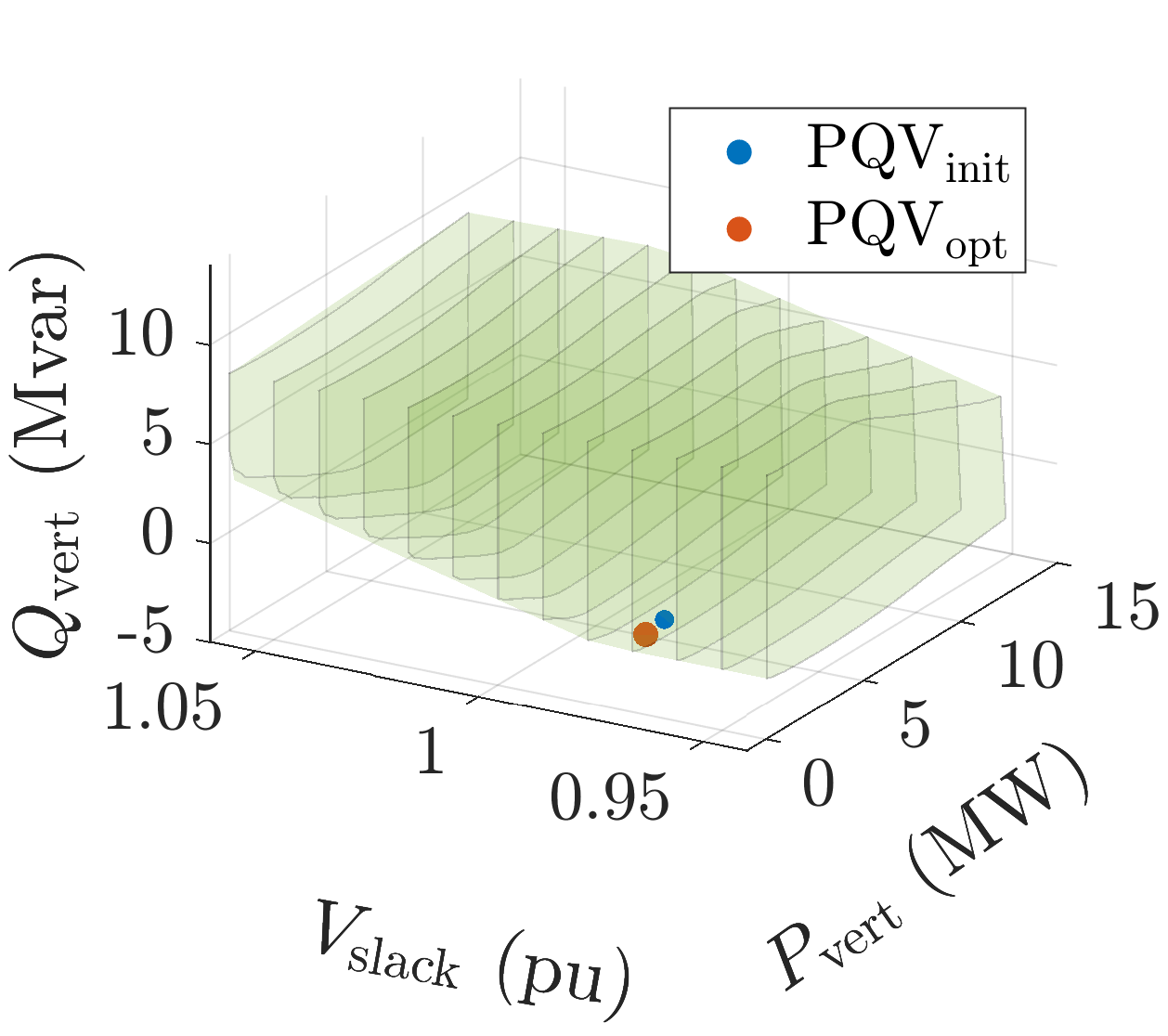}%
\label{fig_first_case}}
\hfil
\subfloat[]{\includegraphics[width=1.25in]{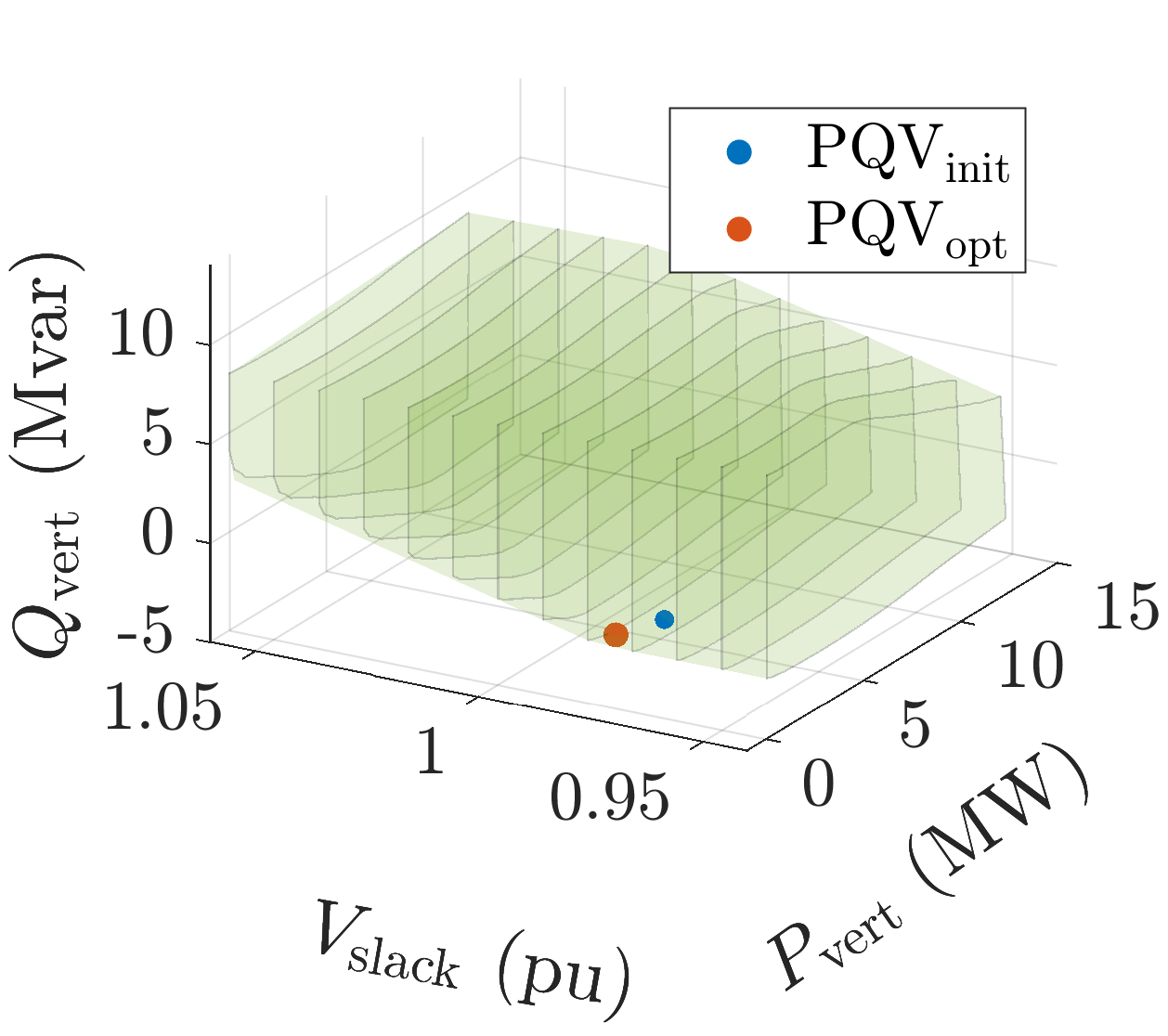}%
\label{fig_second_case}}
\hfil
\subfloat[]{\includegraphics[width=1.25in]{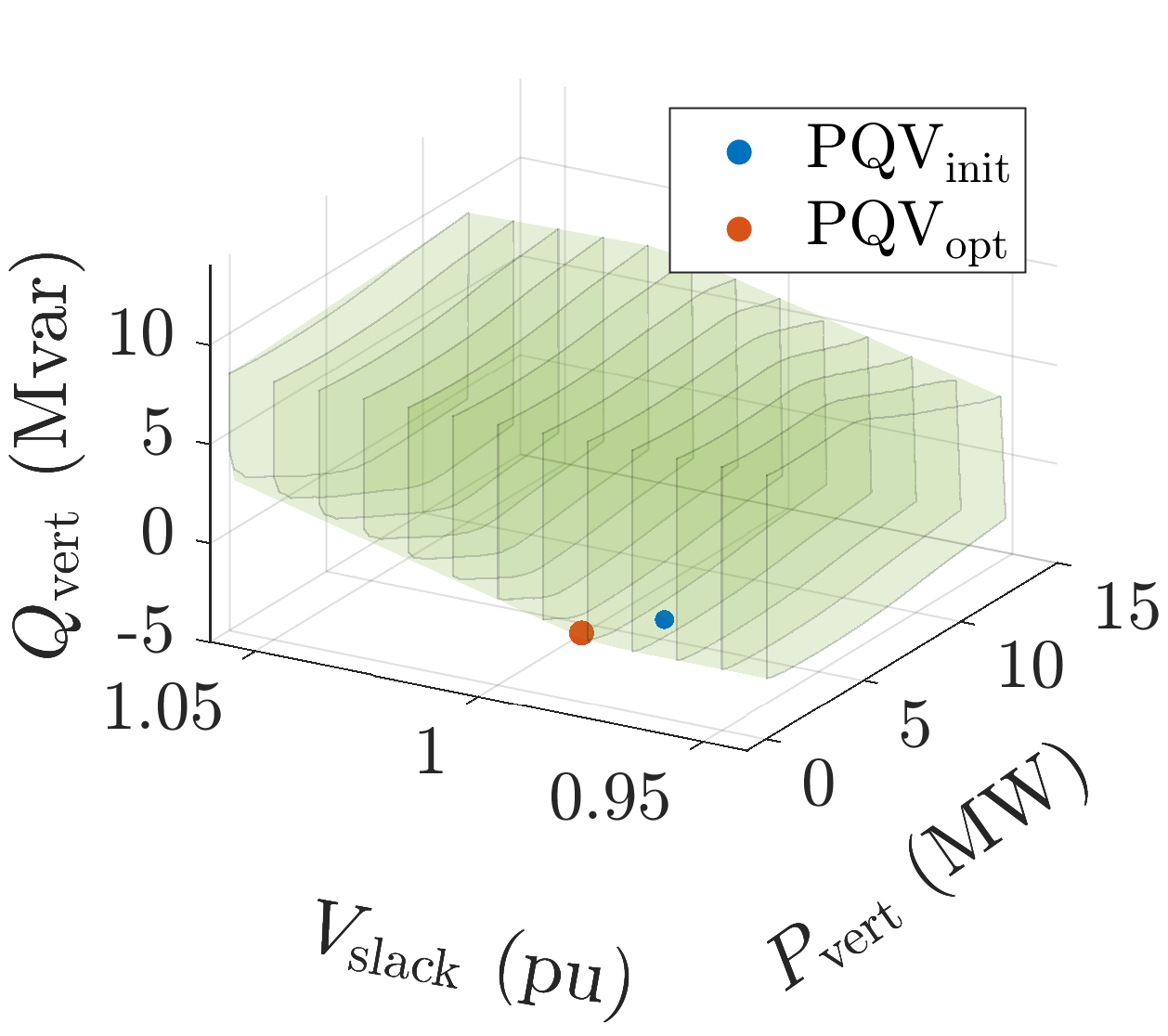}%
\label{fig_second_case}}
\hfil
\subfloat[]{\includegraphics[width=1.25in]{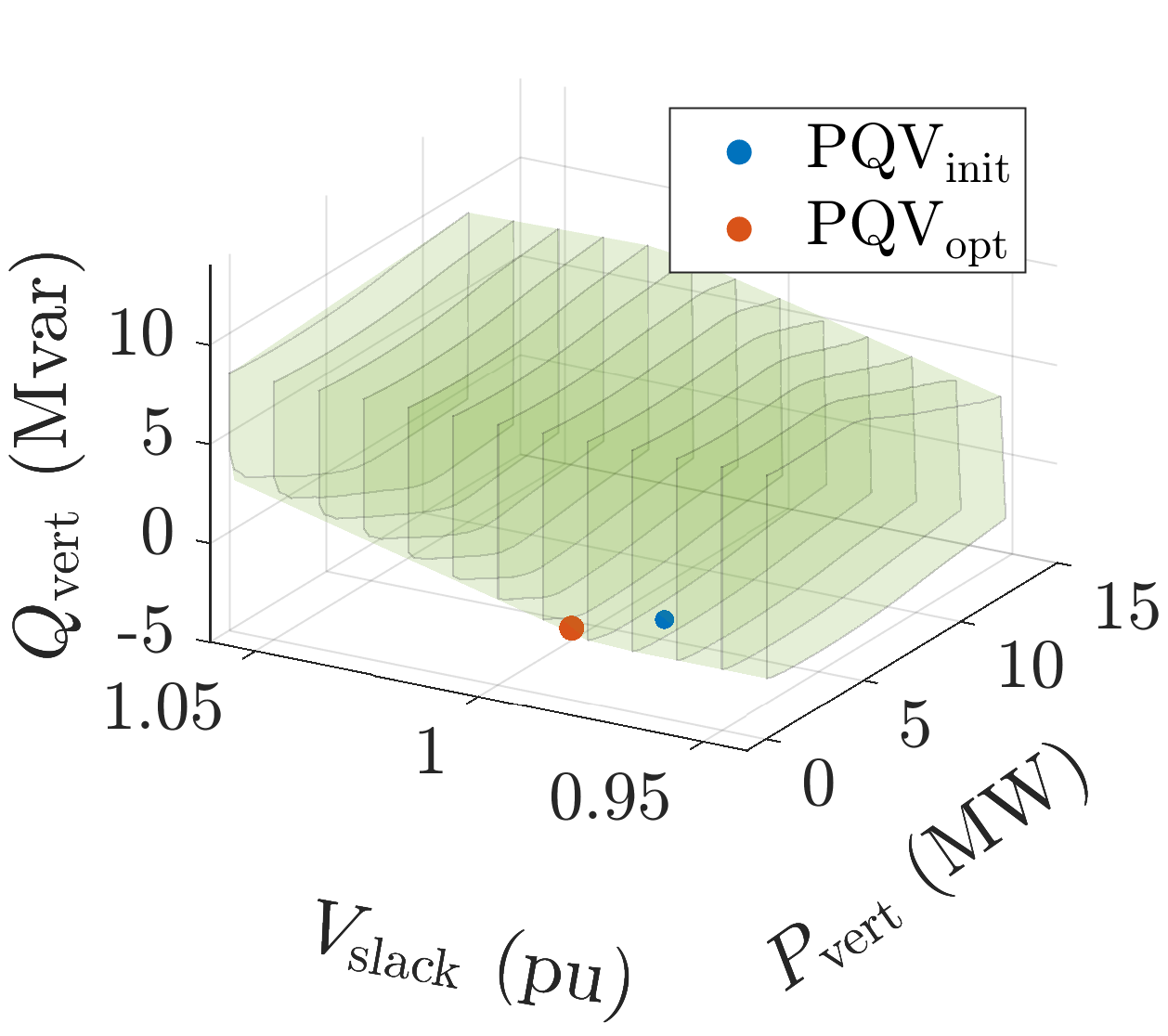}%
\label{fig_second_case}}
\caption{Deviation of the PQ(V) operating point at bus id=20 in response to increased deviations due to reactive power injection: (a) $Q_\mathrm{bus \ id,29}=-50 \ \mathrm{Mvar}$ (b) $Q_\mathrm{bus \ id,29}=-100 \ \mathrm{Mvar}$ (c) $Q_\mathrm{bus \ id,29}=-150 \ \mathrm{Mvar}$ (d) $Q_\mathrm{bus \ id,29}=-200 \ \mathrm{Mvar}$}
\label{fig:fig_Qbigbus20}
\end{figure*}

Further simulation results are obtained for increased reactive power injection at a particular bus. The deviation of PQ(V) operating point from the MV grid aggregated 3D PQ(V) FORs are investigated. The MILP based approach arrives at permissible solutions, however, the optimization costs in the objective function require adaptation for varying scenarios. The underlying cause requires further investigation and a uniform method is still to be determined. Therefore, the results are demonstrated with the convexification based approach in Fig. \ref{fig:fig_Qbigbus7}, \ref{fig:fig_Qbigbus20}. Observations depict a displacement of the optimized operating point  at buses 7, 20 from the initial towards incrementally increasing bus voltages. The polyhedral constraints presented by the 3D PQ(V)-FORs are not violated, indicating robustness of the approaches to the increased deviations in grid power injections. However, in case violations do exist, further iterations are required to correct the operating point within the 3D PQ(V)-FOR bounds. In practice, a suitable underestimation of the FOR volume can be sufficient to arrive at desirable results without the requirement for further iterations.

\section{Conclusion and Outlook}
The MV active and reactive power flexibilitiy potential (PQ-capability or FORs) are aggregated at the HV/MV interconnection and offered to the HV grid operator as operational flexibilities. The two dimensional FORs are further extended considering the effect of HV bus voltages on the corresponding underlying MV grid flexibility potentials. The FORs are accordingly enhanced with the HV bus voltage as a third dimension resulting in 3D PQ(V)-FORs. Subsequent usage of the 3D PQ(V)-FORs to relieve representative congestions and voltage problem at the HV grid is demonstrated using two different strategies. The first employs a MILP based formulation. The polyhedron is segmented into convex sections, within which an optimal solution is determined. The second strategy based on linear programming utilizes a convex hull approximation of the 3D PQ(V)-FOR. Therefore, the enclosed space is determined as an intersection of half-spaces resulting in planar equations, subsequently integrated into the linear optimization based HV grid operational management. The optimal solutions are determined at computationally fast speeds of $\leq$10 seconds. The results for the MILP and convexification based approaches are demonstrated for a representative scenario with reliable solutions in both cases, however, the convexification based approach is computationally faster. The solutions are consistent and the speed presents advantages for solving the problem in short term planning processes. The robustness is further demonstrated for increased perturbance, for example, increased reactive power injections from installed static compensators at the HV grid level. 

In future researches, different methods e.g., for day ahead and intraday planning, heuristic based methods can be examined, as they can potentially offer improved solutions, however, at the cost of significantly increased computation times. Furthermore, the volatility of the renewables requires consideration of uncertainties and consequently adequate security margins for flexibility acquisition from the FORs. Therefore, integration of uncertainties in the HV grid operational management method using 3D PQ(V)-FORs requires to be developed considering forecast time horizons. Subsequently, apart from the technical feasibilities, economic analyses and remuneration schemes are required to monetarize the methods for implementation in a flexibility market. 

 
\bibliographystyle{ieeetr}
\bibliography{Aggregated distribution grid flexibilities in subtransmission grid operational management}

\section*{Biography Section}
\vspace{4pt}
\begin{IEEEbiography}[{\includegraphics[width=1in,height=1.25in,clip,keepaspectratio]{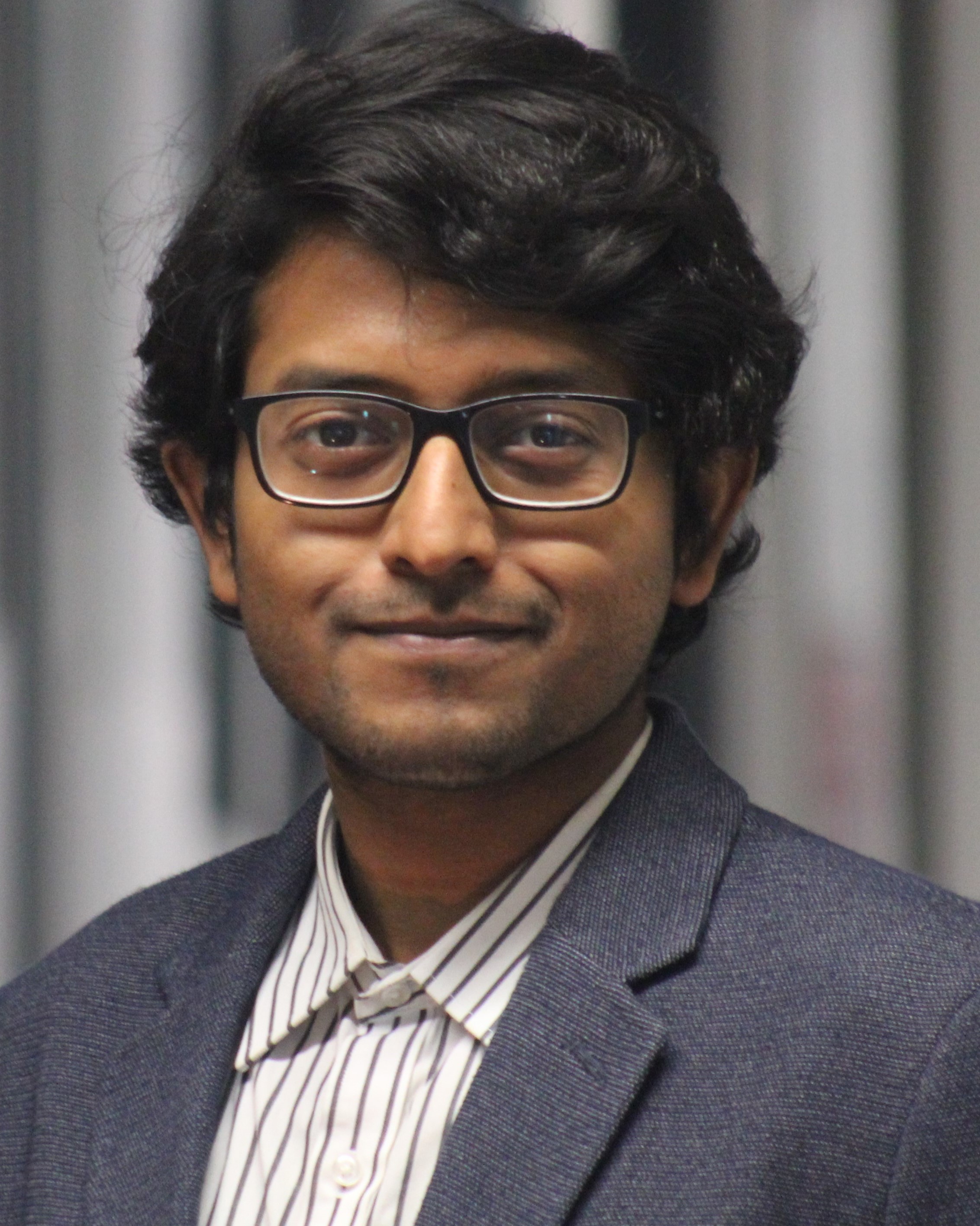}}]{Majumdar, Neelotpal} was born in Kolkata, West Bengal, India in 1993. He received his B.Tech.in Electrical Engineering (2016) from Maulana Abul Kalam Azad University of Technology (formerly West Bengal University of Technology) and M.Sc. degree in Electrical Power Engineering from the RWTH Aachen University, Aachen, Germany, in 2020. Since 2020, he has been working towards a PhD. in Electrical Engineering from the Institute of Electric Power Systems, Leibniz Universität Hannover. His research interests include optimized ancillary services provision between multiple voltage levels and optimal operation under consideration of power system uncertainties.
\end{IEEEbiography}
\begin{IEEEbiography}[{\includegraphics[width=1in,height=1.25in,clip,keepaspectratio]{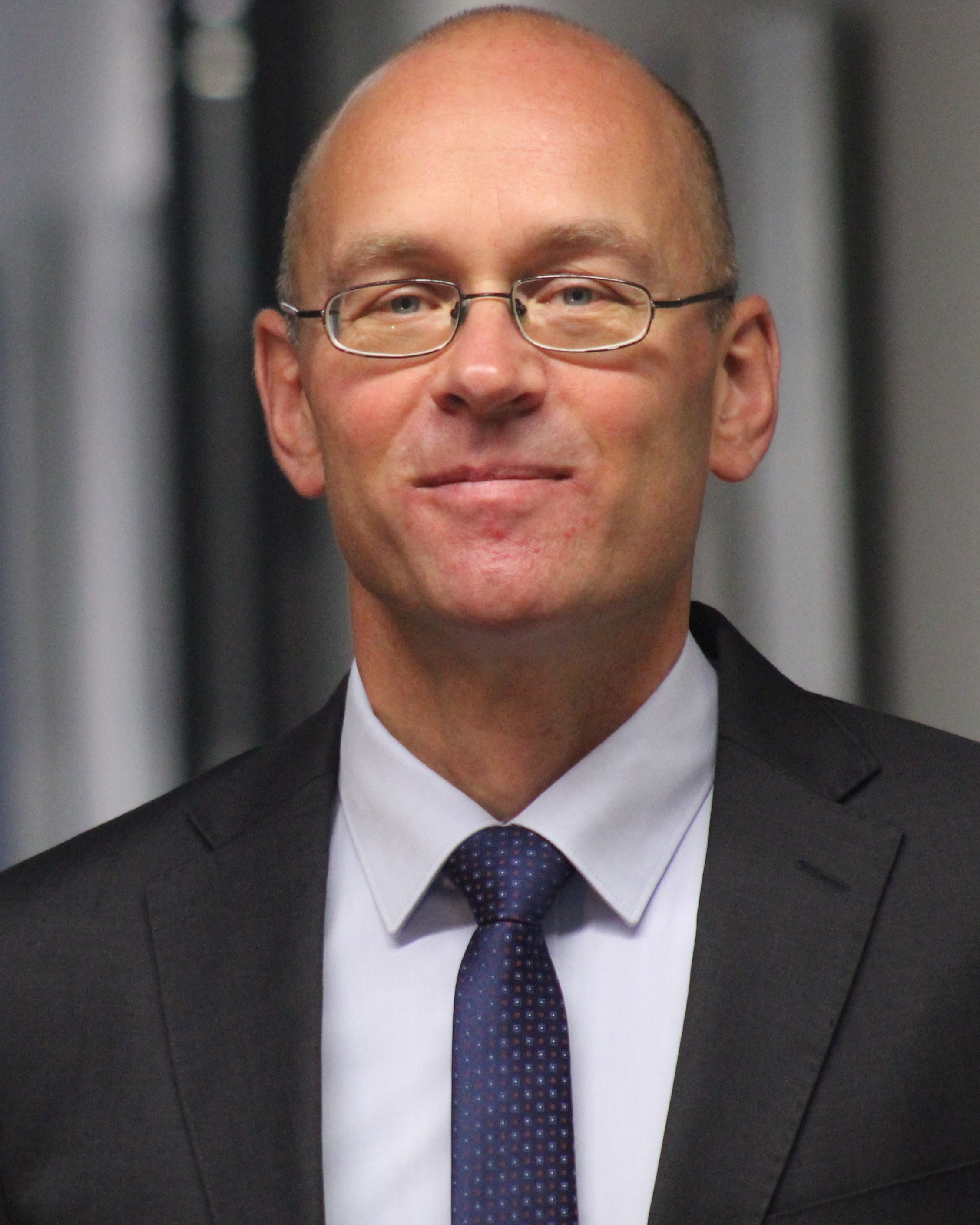}}]{Hofmann, Lutz}was born in Bad Oeynhausen, Germany in 1968. He received the Dipl.-Ing. and Dr.-Ing. degrees from the Leibniz Universität Hannover, Hanover,Germany, in 1994 and 1997, respectively. In 2002, he concluded his professorial dissertation in electric power engineering. In 2002 and 2003,he was a Project Manager for the engineering and consultant company Fichtner in Stuttgart,Germany. From 2004 to 2007, he was with the German transmission system operator E.ON Netz GmbH, Bayreuth, Germany, in the Network Planning Department.Since 2007, he has been a Full Professor and the Head of the Institute of Electric Power Systems, Leibniz Universität Hannover. Since 2011, he has also been the Head of the Department of Transmission Grids, Fraunhofer IWES, Kassel, Germany. His research interests are modeling and simulation of electric power systems, integration of renewable and decentralized energy sources, and power quality.
\end{IEEEbiography}
\vspace{4pt}
\vfill
\end{document}